\RequirePackage{fix-cm}
\RequirePackage{amsmath}
\documentclass[smallextended]{svjour3}       % onecolumn (second format)
\smartqed  % flush right qed marks, e.g. at end of proof
\usepackage{graphicx}
\usepackage[utf8x]{inputenc}
\usepackage{amsmath}
\usepackage{latexsym}
\usepackage{amssymb}
\usepackage{braket}
\usepackage{xcolor}
\usepackage{hyperref}
\hypersetup{colorlinks,linkcolor=blue,linktoc=page,citecolor=blue,filecolor=blue,urlcolor=blue}
\usepackage{url}
\usepackage[numbers,sort&compress]{natbib}
\bibpunct{(}{)}{,}{a}{}{,}
\usepackage{threeparttable}
\usepackage{comment}
\usepackage{booktabs}
\usepackage{enumitem}
\usepackage{siunitx}

\newcommand{\xmm}{\textit{XMM-Newton}}
\newcommand{\erosita}{\textit{eROSITA}}
\newcommand{\athena}{\textit{Athena}}

\begin{document}

\title{Scattering efficiencies measurements of soft protons at grazing incidence from an \textit{Athena} Silicon Pore Optics sample}



\titlerunning{Scattering efficiencies of soft protons from \textit{Athena} SPO}        % if too long for running head

\author{Roberta Amato \and Sebastian Diebold \and Alejandro Guzman \and Emanuele Perinati \and Chris Tenzer \and Andrea Santangelo \and Teresa Mineo}


\institute{Roberta Amato \at
              Dipartimento di Fisica e Chimica - Emilio Segr\'e, Università degli Studi di Palermo, via Archirafi, 36, 90123 Palermo, Italy \\
              INAF-IASF Palermo, via Ugo La Malfa, 153, 90146 Palermo, Italy \\
              IAAT, University of T\"ubingen, Sand 1, 72076 Tübingen, Germany\\
              \email{astro.roberta@gmail.com}           %  \\
           \and Sebastian Diebold \and Alejandro Guzman \and Emanuele Perinati \and Chris Tenzer \and Andrea Santangelo \at IAAT, University of T\"ubingen, Sand 1, 72076 Tübingen, Germany
           \and Teresa Mineo \at INAF-IASF Palermo, via Ugo La Malfa, 153, 90146 Palermo, Italy
           }

\date{Received: date / Accepted: date}

\maketitle

\begin{abstract}
Soft protons are a potential threat for X-ray missions using grazing incidence optics, as once focused onto the detectors they can contribute to increase the background and possibly induce radiation damage as well. The assessment of these undesired effects is especially relevant for the future ESA X-ray mission \textit{Athena}, due to its large collecting area. To prevent degradation of the instrumental performance, which ultimately could compromise some of the scientific goals of the mission, the adoption of ad-hoc magnetic diverters is envisaged. Dedicated laboratory measurements are fundamental to understand the mechanisms of proton forward scattering, validate the application of the existing physical models to the \textit{Athena} case and support the design of the diverters. In this paper we report on scattering efficiency measurements of soft protons impinging at grazing incidence onto a Silicon Pore Optics sample, conducted in the framework of the EXACRAD project. Measurements were taken at two different energies, $\sim$470 keV and $\sim$170 keV, and at four different scattering angles between 0.6° and 1.2°. The results are generally consistent with previous measurements conducted on \textit{eROSITA} mirror samples, and as expected the peak of the scattering efficiency is found around the angle of specular reflection.

\end{abstract}

\section{Introduction}
\label{section:introduction}

Orbiting X-ray observatories are subjected to the impact of charged particles of different origins, e.g. cosmic rays, solar wind and solar events, electrons and protons trapped in the Earth magnetosphere, etc. Amongst them, protons with typical energies ranging from a few tens to a few hundreds of kiloelectronvolts, the so-called ‘soft protons' (SP), are scattered and funneled towards the focal plane when impinging at grazing incidence on X-ray optics. Once they reach the instruments at the focal plane, they significantly contribute to the level of non-X-ray background (NXB), and potentially damage the detectors in the most severe cases. The effects of SP were first observed on the \textit{Chandra X-ray Observatory} \citep{Weisskopf2000} and on \xmm{} \citep{Jansen2001}, both launched in 1999. To this day SP affect the operability of these X-ray missions by significantly reducing their useful observing time and their duty cycles -- for instance, the observing time of \xmm{} is reduced by ∼30–40\,\% \citep{XMMbkg-3}.

SP constitute an even bigger issue for \athena{} \citep[\textit{Advanced Telescope for High Energy Astrophysics},][]{Nandra2013}, an L-class ESA mission, planned to be launched in the early 2030s. The telescope will be equipped with Silicon Pore Optics (SPO) \citep{Willingale2013}, wafers of silicon coated with high-Z materials, stacked one upon the other and arranged to cover the whole pupil of the telescope.

\athena{} will orbit around the second Lagrangian point L2\footnote{Currently, there are strong suggestions in favour of an L1 orbit, where the particle environment is better known, though highly variable} of the Earth-Sun system, at a distance of about 1.5 million kilometer from Earth in the anti-Sun direction. This orbit lies in the tail of the Earth magnetosphere, where the particle environment is extremely variable \citep{Lotti2017}. At present, the science driven requirement for SP is that their flux at the focal plane has to be  $<5\times10^{-4}$ cts s$^{-1}$ cm$^{-2}$ keV$^{-1}$ (corresponding to 10\,\% of the total NXB), in the 2--10\,keV energy band, for 90\,\% of the observing time\footnote{\url{https://www.cosmos.esa.int/documents/400752/507693/Athena_SciRd_iss1v5.pdf}}.
In order to have a thorough estimate of the SP flux expected at the focal plane, knowing the scattering efficiency of soft protons at grazing incidence from \athena{}'s SPO is fundamental. To achieve this goal, we performed experimental scattering efficiency measurements with a SPO sample.

Similar experimental measurements have already been performed for \xmm{} \citep{Rasmussen} and \erosita{} \citep{Diebold,SPIE2017} mirror samples. Especially the measurements on \erosita{} showed that the scattering is always non-elastic, with energy losses of the order of a few tens of kiloelectronvolts, depending on the energy of the incident beam. The scattering efficiency always peaks close to the angle of specular reflection and is higher for lower incident angles, while it is almost independent of the energy of the impinging protons.

A model for the scattering efficiency of ions at grazing incidence from the surface of a material  was developed by \cite{Remizovich}, in non-elastic approximation. However, the model is generally valid for any surface and contains some approximations, therefore a simple application of the Remizovich formula to the \erosita{} data did not exactly reproduce the efficiencies measured in the laboratory \citep{Amato2020}. A fit of the experimental data with the Remizovich formula led to the evaluation of the parameter enclosing the micro-physics of the scattering, so that a new analytical semi-empirical model that better reproduces the data from the \erosita{} mirror sample was derived \citep{Amato2020}. This model can be used to assess the SP flux expected at the instrumental focal plane of the satellite. The model can also be applied to \athena{}, provided that the proton scattering properties of the SPO are experimentally determined.

In this publication, we present the first measurements of scattering efficiencies of low-energy protons off a SPO sample. These experimental activities were conducted in the framework of the EXACRAD (Experimental Evaluation of Athena
Charged Particle Background from Secondary Radiation and Scattering in Optics) project funded by ESA. The paper is structured as follows: we first describe the laboratory setup and the elements along the beam line in Sect.\,\ref{section:setup}; in  Sect.\,\ref{sec:experimental_efficiency_definition} we illustrate how the scattering efficiency is derived from the raw data; the new data are presented in Sect.\,\ref{section:results} and are compared to the data from \erosita{} as well as to the semi-empirical model mentioned above in Sect.\,\ref{fig:comparison_SPO_erosita}; finally, we draw our conclusions in Sect.\,\ref{section:conclusions}.

\section{Experimental setup}
\label{section:setup}

\begin{figure}[htbp]
\centering
\includegraphics[width=1.\textwidth]{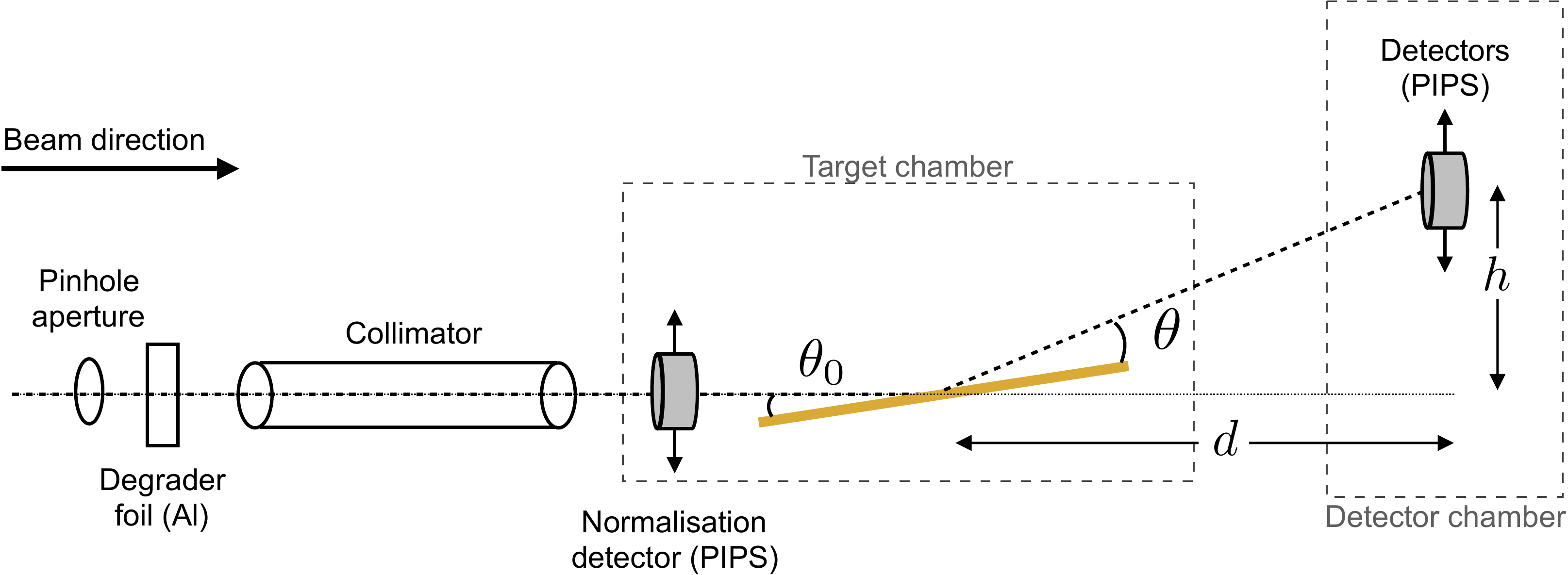}
\caption{Schematic drawing (not in scale) of the beamline setup. The proton beam enters the setup from the left-hand side. It encounters the pinhole aperture (1 mm in diameter), the Al degrader foil (0.002 mm thick) and the collimator. Inside the target chamber, the normalisation detector can be lowered down to intercept the beam for the normalisation measurements. If the normalisation detector is not in the line of the beam, then protons are reflected from the SPO sample (in yellow) towards the detector chamber, where they hit the central and lateral detectors. The incident angle $\theta_0$ between the line of the beam and the mirror varies with the inclination of the target plate, while the scattering angle $\theta$ between the mirror and the detectors in the detector chamber varies with the their height $h$. The distance $d$ between the target plate and the vertical ax of the central detectors is fixed to 942 mm.}
\label{fig:beam_line_scheme}
\end{figure}

\begin{figure}[htbp]
\centering
\includegraphics[width=1.\textwidth]{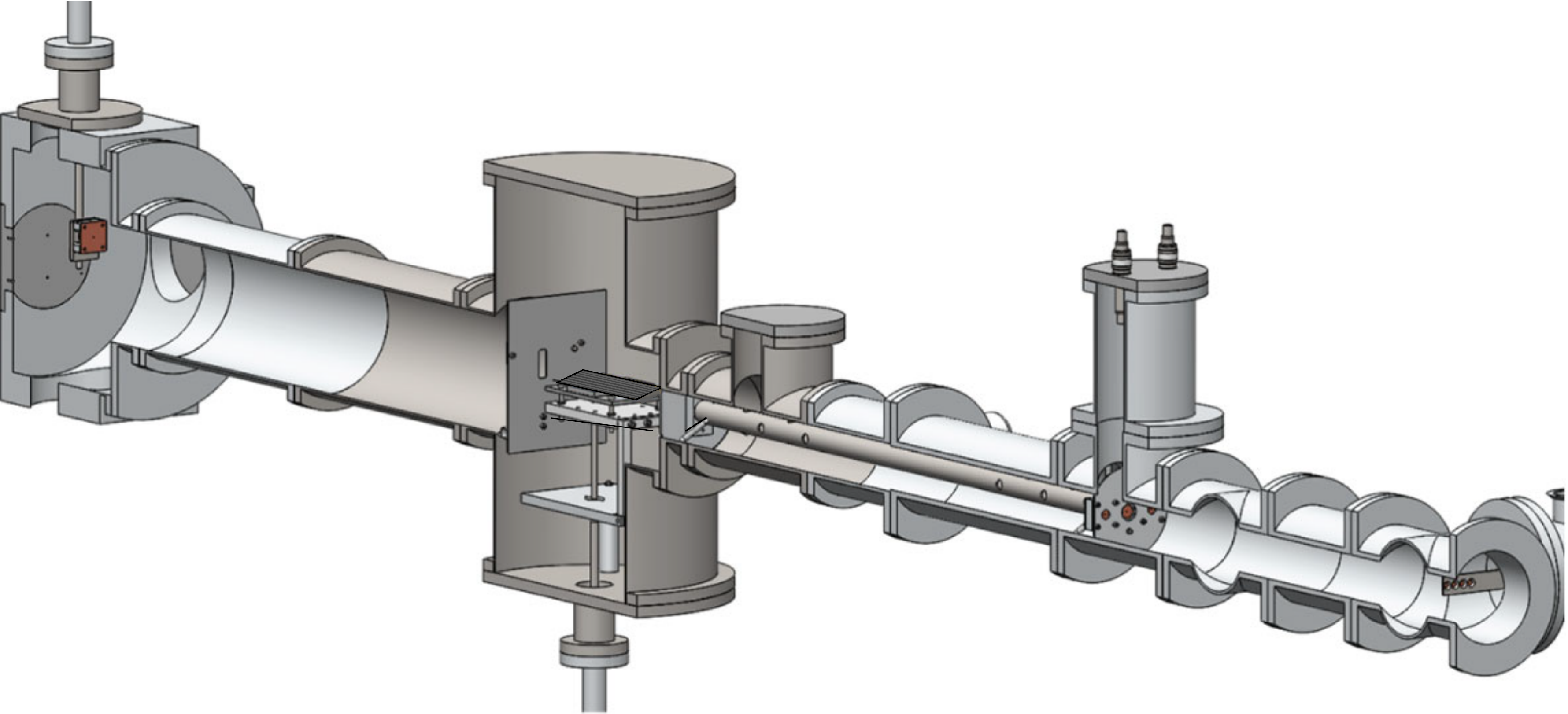}
\caption{A CAD model of the beamline \citep[same as][]{Diebold}. The proton beam enters the setup from the right and moves towards the left. The SPO sample is located in the target chamber, while the detector is placed in the chamber at the end of the beamline (detector chamber). A second detector (not shown in the picture) was placed next to the central one, with an angular distance of $\sim2^\circ$.}
\label{fig:beam_line_CAD}
\end{figure}

\begin{figure}[htbp]
\centering
\includegraphics[width=1.\textwidth]{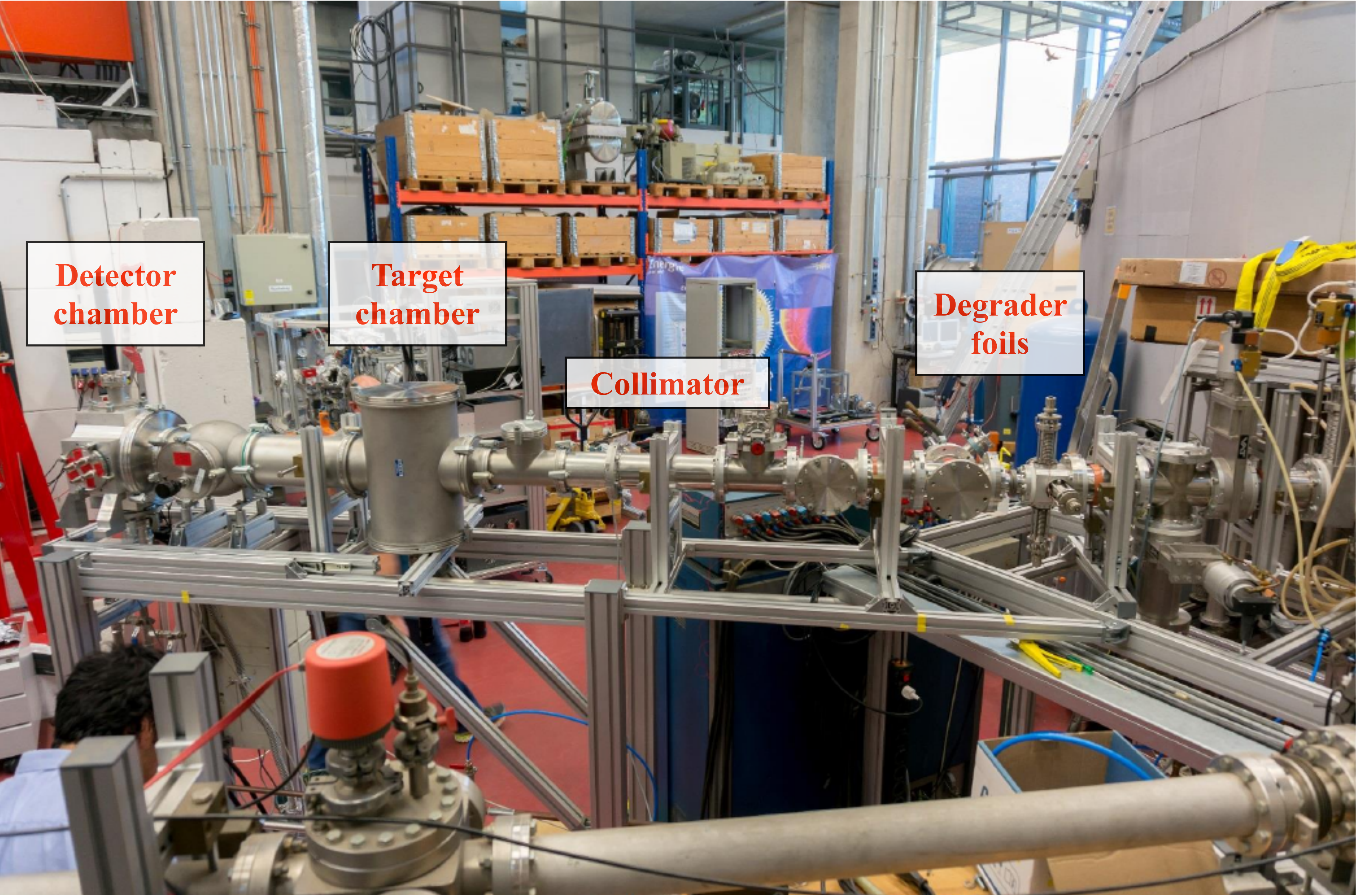}
\caption{Picture of the beamline at the Van der Graaff accelerator facility. The beam direction is from the right to the left. The position of the degrader foils and the collimator inside the beamline are pointed out, as well as the detector and target chambers.}
\label{fig:beamline_pic}
\end{figure}

The experiment was conducted at the 2.5 MV Van de Graaff accelerator at the Goethe University (Riedberg Campus) in Frankfurt am Main. The setup of the beamline, similar to that of \cite{Diebold,SPIE2017}, is given in Figs.\,\ref{fig:beam_line_scheme}, \ref{fig:beam_line_CAD}, and \ref{fig:beamline_pic}.

\begin{figure}[htbp]
\centering
\includegraphics[width=.45\textwidth,height=.18\textheight]{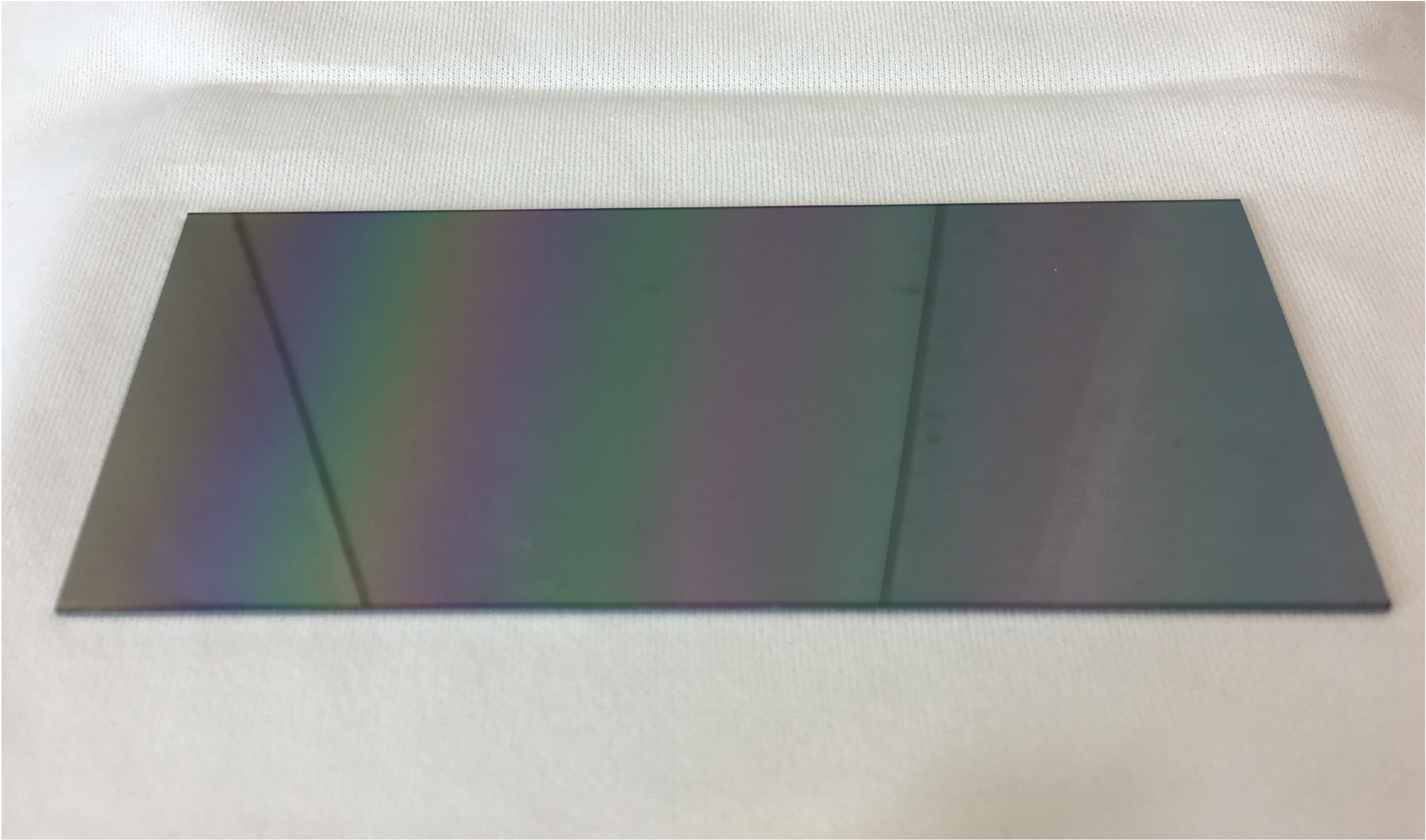}\quad
\includegraphics[width=.45\textwidth, height=.18\textheight]{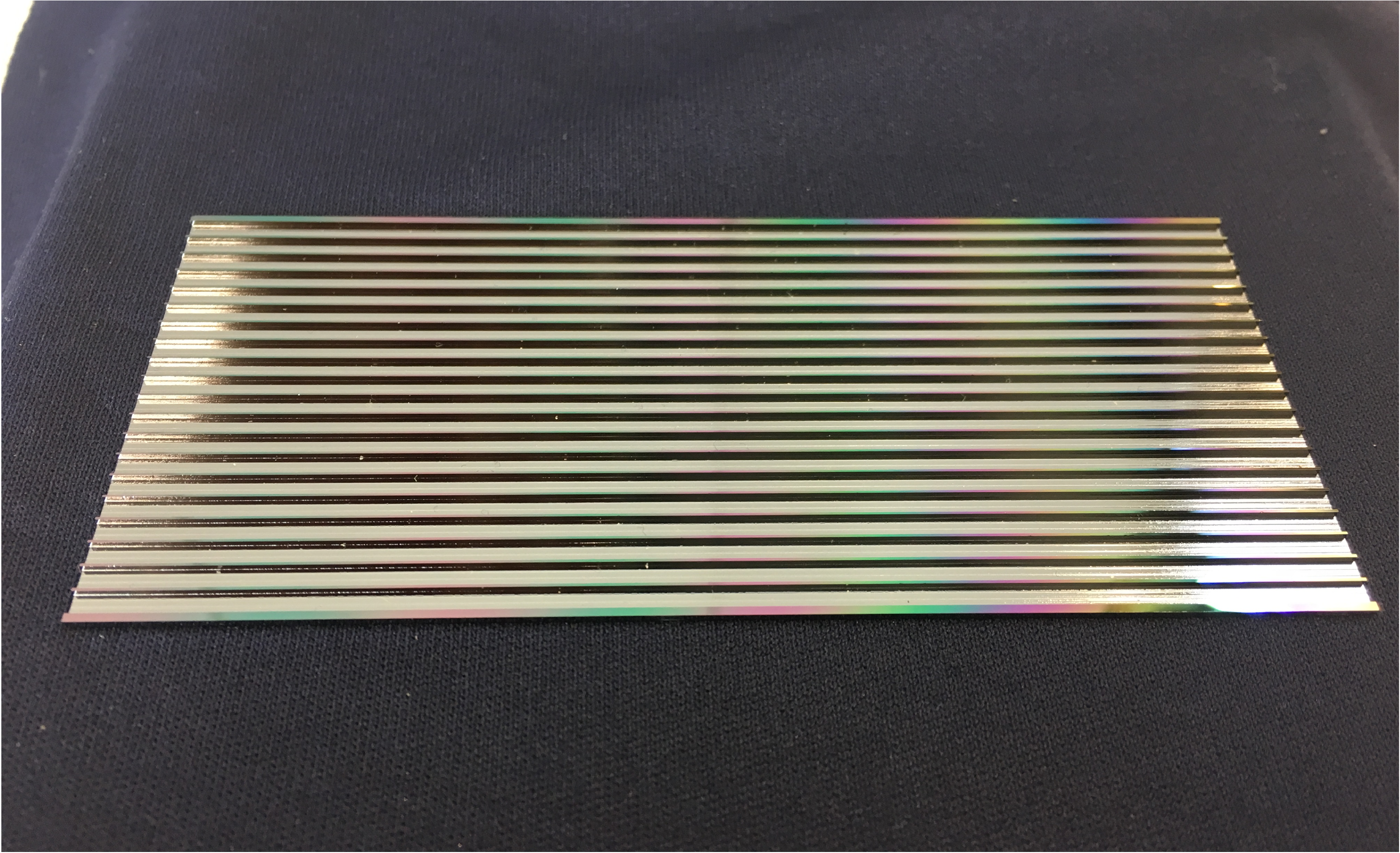}
\caption{Pictures of the SPO sample, with the reflecting surface up (left) and down (right). }
\label{fig:mirros_pics}
\end{figure}

\subsection{Beamline setup}
\label{subsec:beamline_setup}

Protons enter the beamline through a copper pinhole aperture of the diameter of 1 mm, which reduces the size of the incoming beam to prevent pile-up and to maintain reasonable rates on the detectors. Successively, the beam goes through a 0.002 mm thick aluminium foil, which degrades the incoming beam energy below the lower limit of the accelerator. The degraded beam enters, at this point, a 78 cm long collimator, which directs part of the widened beam directly to the target. Two further apertures are positioned at the entrance and at the exit of the collimator, both with a diameter of 1 mm. This diameter limits the smallest possible incident angle to $\sim$0.5$^\circ$. The apertures are supported in their position by 2 mm aluminum plates, which absorb any proton of the degraded beam not entering the apertures and being scattered by the inner walls of the collimator and of the beamline itself. 

The SPO target (Fig.\,\ref{fig:mirros_pics}), provided by \textit{cosine}\footnote{\url{https://www.cosine.nl/cases/silicon-pore-optics-mirror-modules-spo-for-astronomy/}.}, consists of a 110 mm long single silicon wafer, 0.775 mm thick, grooved in the bottom, and coated on top with a 10 nm of iridium and 7 nm of silicon carbide\footnote{Though iridium is the 
baseline coating material, a low-Z overcoating is also considered to 
improve the reflectivity at lower energies. Different low-Z materials 
and thicknesses are currently under investigation.}.
It is located in an apposite chamber (hereafter called target chamber) and mounted on a tiltable plate. The height of the target can be adjusted by a set of screws underneath the plate. A linear manipulator is used to change the inclination of the plate, i.e., of the incident angle ($\theta_0$). The pivoting point is is several centimeters below the line of the beam, so that the target can be completely removed from the beam, allowing for a determination of the primary beam position on the detector plane. The manipulator is set below the target chamber and hence can be easily accessed when the system is under vacuum. 

Between the exit of the collimator and the target plate, a Passivated Implanted Planar Silicon (PIPS) detector\footnote{The PIPS detectors used in this experiment have a nominal depletion region of 0.1 mm and a lower energy threshold of a few tens of keV.} is mounted on a push-pull manipulator, at the same height of the beamline. This detector is used to register the amount of flux of the incident beam impinging on the target, useful to have normalisation measurements. This detector will be called hereafter 'normalisation detector'. The push-pull manipulator permits a fast removal of the detector, guaranteeing a measure of the impinging proton flux ($\Phi_{inc}$, cfr.\,Eq.\,\ref{eq:lab_efficiency}) for each measure of the scattered beam (see Sect.\,\ref{sec:experimental_efficiency_definition} for the need of having frequent normalisation measurements). An aluminium blind with an aperture of 3 mm is set on top of the normalisation detector to avoid saturation. Lastly, downstream of the target chamber, a thick aluminum sheet, with a slit of 3 cm height and 1 cm width, is installed a few centimeters after the target plate. This window lets pass only the protons on the line of the beam, while the sheet absorbs all the ones that have been scattered by the inner walls or by the other elements in the target chamber. 

At the end of the beamline, a second chamber (hereafter detector chamber) hosts two more PIPS detectors, called ‘central detector' and ‘lateral detector', respectively, used to register the on-axis and off-axis fluxes ($\Phi_\text{scat}(\theta_0,\theta,\phi)$, cfr.\, Eq.\,\ref{eq:lab_efficiency}) of the beam scattered by the target. They are mounted on a second linear manipulator, which allows a spatially resolved sampling of the scattered beam. The distance between the center of the target plate and the detection plane is 942 mm. The central detector is aligned with the beam, while the lateral detector is set on the left. This configuration allows for a coverage of the scattered beam on the the incident direction and at an azimuthal angle $\phi$ of $1.97^\circ\pm0.13^\circ$. On top of each detector there is a blind with an aperture of a diameter of 1 mm for the central detectors and of 3 mm for the lateral detector, respectively. They reduce the solid angle of the detectors with respect to the mirror center to about $8\times10^{-7}$ sr and $2\times10^{-5}$ sr for the central and lateral detector, respectively. % (cfr.\,Eq.\,\ref{eq:solid_angle}, Sect.\,\ref{sec:experimental_efficiency_definition}). 

\subsection{Data acquisition chain}
\label{subsec:data_acquisition_chain}

\begin{figure}[htbp]
\centering
\includegraphics[scale=0.3]{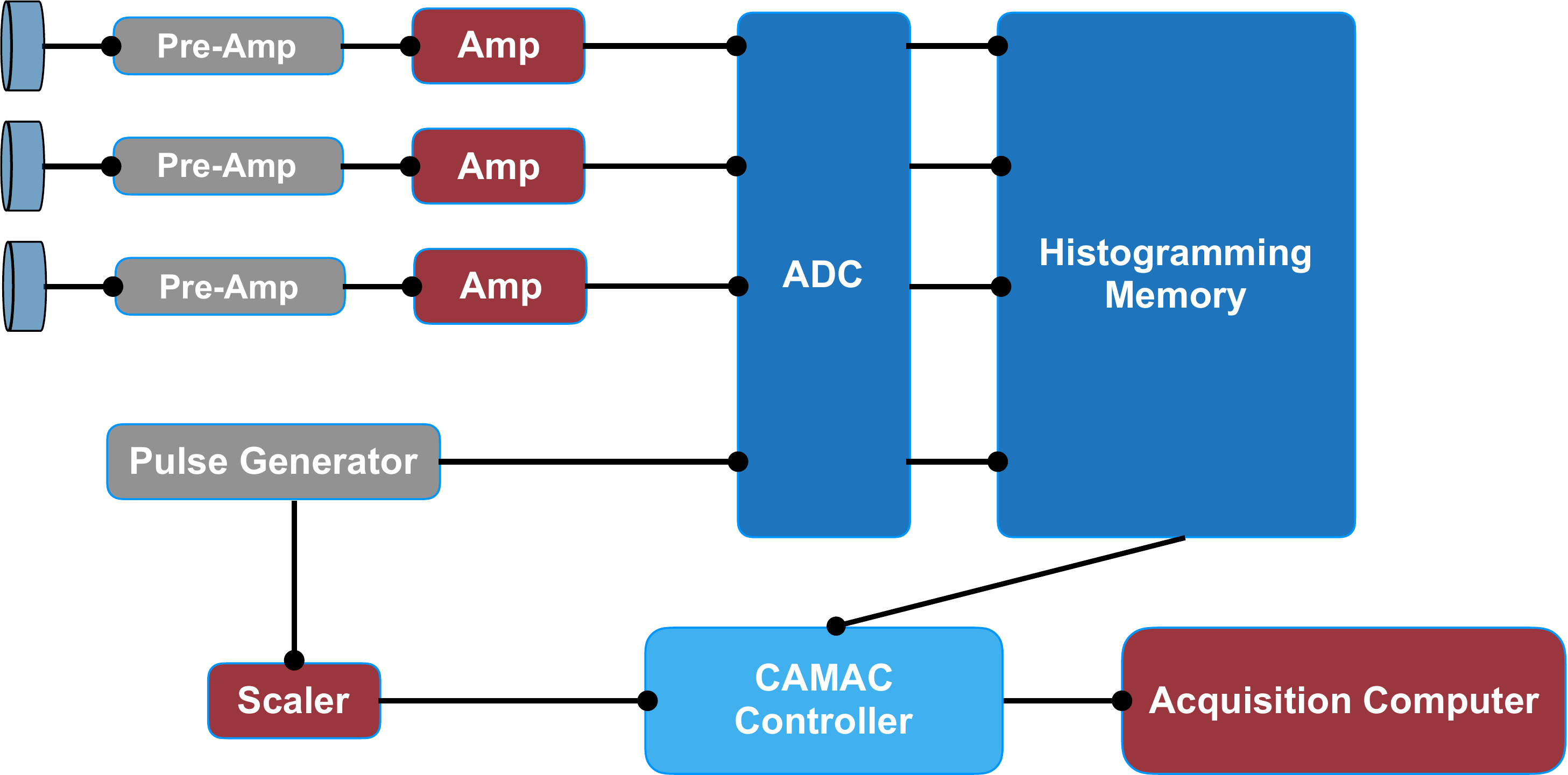}
\caption{Data flow of the electronic chain for the acquisition of the experimental data. The analogical signal from the PIPS detectors first goes through a pre-amplifier and an amplifier, then it is converted into a digital signal by the ADC, and finally it is stored in the histogramming memory. Contemporary, a pulse genarator sends signal at a fixed frequency to the ADC and to a scaler. The digitised signals are read out by a CAMAC controller unit, which transmits them to a computer once the measurement is finished.}
\label{fig:acquisition_chain}
\end{figure}

The pulse signal produced by the PIPS is amplified and digitalised trough several analogical/digital electronic components. A flow chart is given in Fig.\,\ref{fig:acquisition_chain}. 

The PIPS detectors produce a pulse with an amplitude proportional to the energy of the incident particle. The pulse signal from each PIPS goes through its own pre-amplifier and amplifier and it is then  digitalised by the Analog to Digital Converter (ADC). The ADC receive the continuous signal (from 0 to $\sim$10 V) from the three channels -- one for each detector -- and convert them into discrete signals, distributing it into 8192 bins, with a resolution of 1.22 mV. The digitised signals are then passed to the histogramming memory, which produces an histogram for each channel. Once the measurement is done, the histograms are read out by the CAMAC module and are transferred to a computer, which acquires and stores them as raw data files.

The process of digitalisation of the data within the ADC takes a certain time (fractions of seconds), so that if a new signal comes within that time, it is not registered. To account for this dead-time, a pulse generator, which generates pulses at a fixed frequency, is connected to the ADC and to a scaler, which counts the number of pulses produced by the pulse generator during the acquisition time. The scaler is also fed to the CAMAC control module. The difference between the reading of the counts from the ADC and that from the scaler gives the dead-time correction factor (cfr. Eq.\,\ref{eq:dead_time}). The pulse generator fed to the ADC constitutes another channel, so that the whole acquisition systems consists of four channels, all working simultaneously, and the scaler.  

\subsection{Alignment and angular calibration}
\label{sucsec:alignment_and_angular_calibration}

The alignment of all the movable elements on the beamline, i.e., the pinhole aperture, the slits, the normalisation detector, and the central detector, is done by using a telescope previously aligned with the exit of the accelerator. 

A 520 nm laser %, which can be operated using pulse-width-modulation (PWM), 
is employed to perform the angular calibration. The laser is set right after the pinhole aperture and goes through all the slits. When the target plate is down, the laser reaches the central detector in the detector chamber. In this way, the zero of the beamline, corresponding to $\theta=0^\circ$, can be established. This measurement gives also the vertical offset on the linear manipulator of the central/lateral detectors. 

To calibrate the incident and scattering angles, we use the property of the mirror target to reflect optical light. Hence, we raise the target plate, using its own manipulator, until the light is blocked. Then, we raise the central detector till the laser beam is detected again. Assuming a specular reflection, the angle subtended by the height $h$ of the manipulator will be $\zeta=\theta+\theta_0=2\theta_0$, so that the incident angle can be computed as:
\begin{equation}
    \theta_0=\frac{\zeta}{2}
    \label{eq:incident_angle}
\end{equation}
This operation is repeated several time, so that we end up with different angles corresponding to different readings on the linear manipulator of the target plate. The incident angle can then be determined with a simple linear interpolation. 

\section{Efficiency definition and normalisation measurement}
\label{sec:experimental_efficiency_definition}

\begin{figure}[htbp]
\centering
\includegraphics[width=.5\textwidth]{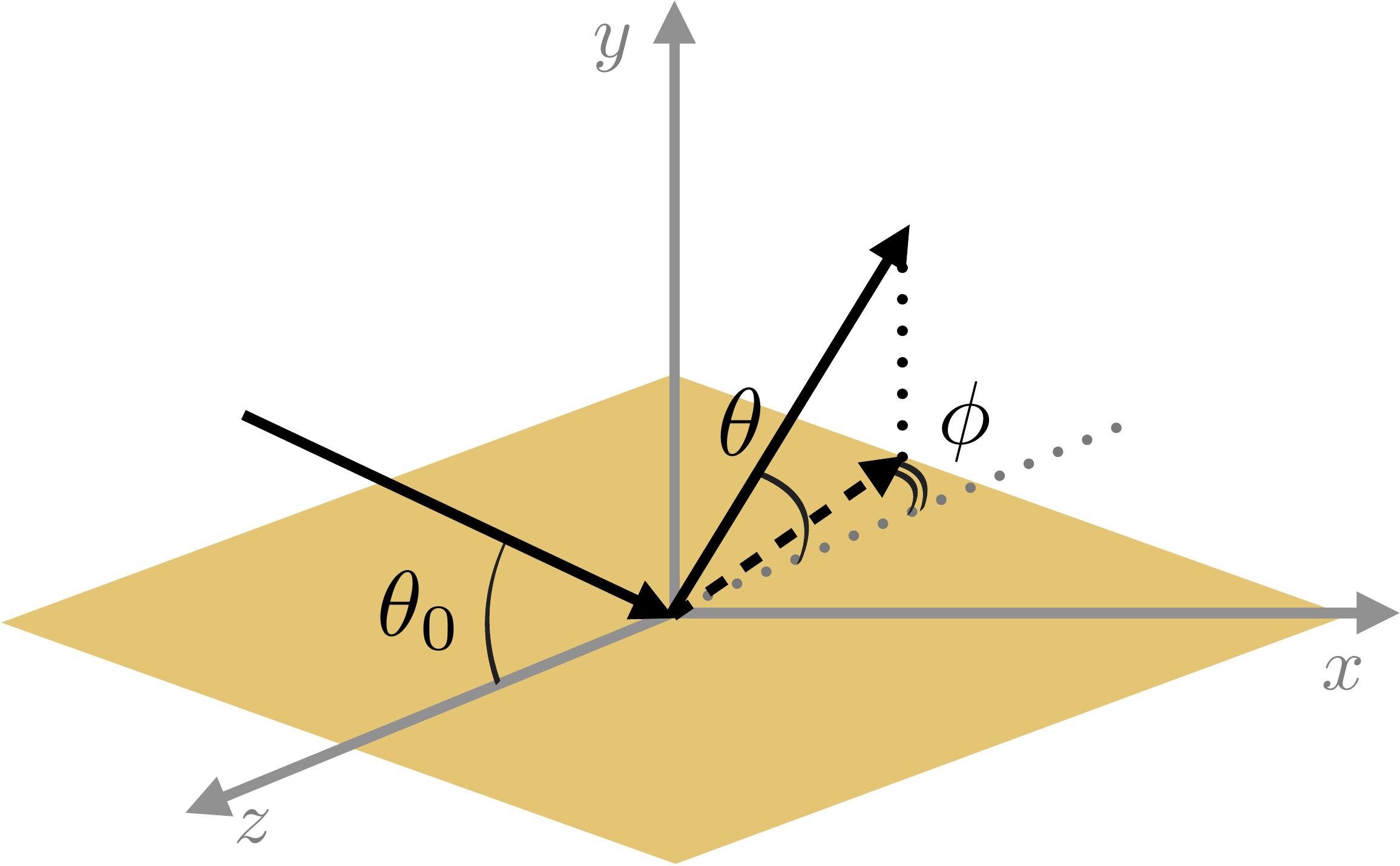}
\caption{Geometrical scheme of the incident plane. The proton beam hits the mirror sample, in the plane $xz$ with an incident angle $\theta_0$ and is pseudo-reflected with a polar scattering angle $\theta$ and an azimuthal scattering angle $\phi$.}
\label{fig:ang_scheme}
\end{figure}

In the laboratory system of reference, the scattering efficiency per unit solid angle can be defined as:
\begin{equation}
    \eta(\theta_0,\theta,\phi)=\frac{1}{\Omega(\theta,\phi)}\frac{\Phi_{\text{scat}}(\theta_0,\theta,\phi)}{\Phi_\text{inc}}
    \label{eq:lab_efficiency}
\end{equation}
where $\theta_0$ is the incident angle, $\theta$ and $\phi$ are the polar and azimuthal scattering angles (see Fig.\,\ref{fig:ang_scheme}), $\Phi_{scat}$ and $\Phi_{inc}$ are the scattered and incident proton count rates, and $\Omega(\theta, \phi)$ is the solid angle subtended by the detector. 

The count rate of the scattered particles is given by the number of  protons $N_\text{scat}$ scattered by the SPO sample reaching the detectors in the detector chamber divided by the integration time $\Delta t_\text{scat}$. In a similar way, the count rate of the incident particles is given by the number of particles $N_\text{inc}$ intercepted by the normalisation detector in front of the mirror chamber divided by the integration time $\Delta t_\text{inc}$. 
The number of counts of incident and scattered protons, $N_\text{inc}$ and $N_\text{scat}$, is obtained by integrating the ADC histograms. This number must be corrected for the dead-time of the ADC (cfr. Sect.\,\ref{subsec:data_acquisition_chain}), so that the effective count rates can be expressed as:
\begin{equation}
    \Phi_\text{scat}(\theta_0,\theta,\phi)=\alpha\,\frac{N_\text{scat}(\theta_0,\theta,\phi)}{\Delta t_\text{scat}} \quad,\quad\quad
    \Phi_\text{inc}=\alpha\,\frac{N_\text{inc}}{\Delta t_\text{inc}}
\label{eq:phi_scat}
\end{equation}
with the correction factor $\alpha$  given by:
\begin{equation}
    \alpha=\frac{N_\text{scaler}}{(N_\text{pulser})_\text{ADC}}
    \label{eq:dead_time}
\end{equation}
where $N_\text{scaler}$ is the number of counts from the pulse generator as read out from the scaler fed to the CAMAC controller module and $(N_\text{pulser})_\text{ADC}$ is the number of pulses from the pulse generator as read out from the ADC (see Fig.\,\ref{fig:acquisition_chain}). 
For an ideal incoming proton beam, the number of incident particles $N_\text{inc}$ is constant in time. However, the beam exiting the Van de Graaff accelerator was not stable, with fluctuations in the direction of the beamline varying in a time range from a few to several tens of minutes. This made necessary to take normalisation measurements before and after each scattering measurement and average them for each scattering data point, so that:
\begin{equation}
    \frac{N_\text{inc}}{\Delta t_\text{inc}}=\frac{1}{2}\left(\frac{N_\mathrm{inc,1}}{\Delta t_\mathrm{inc,1}}+\frac{N_\mathrm{inc,2}}{\Delta t_\mathrm{inc,2}}\right)
\end{equation}
where $N_\mathrm{inc,1}$ and $N_\mathrm{inc,2}$ are the counts in two consecutive normalisation measurements with integration times $\Delta t_\mathrm{inc,1}$ and $\Delta t_\mathrm{inc,2}$, respectively.

Concerning the uncertainties, the one on the scattering angle is given mainly by the errors on the angular calibration, the detector aperture, and the indeterminate position of the impact point of the beam on the mirror surface. The uncertainty on the incident angle $\theta_0$ is dominated by the dimension of the aperture on the central detector and by the length of the target. It resulted in $\sim$\ang{0.1} for all the chosen  scattering angles. Lastly, the uncertainty on the scattering efficiency is mainly given by the intrinsic fluctuation of the proton beam. Minor contributions are due to the count statistics and to the error on the solid angle $\Omega(\theta, \phi)$. The sum of this contributions results in statistical fluctuations of $\pm$20\% on the scattering efficiencies.

\section{Results and discussion}
\label{section:results}

\begin{figure}[htbp]
\centering
\includegraphics[width=1.\textwidth]{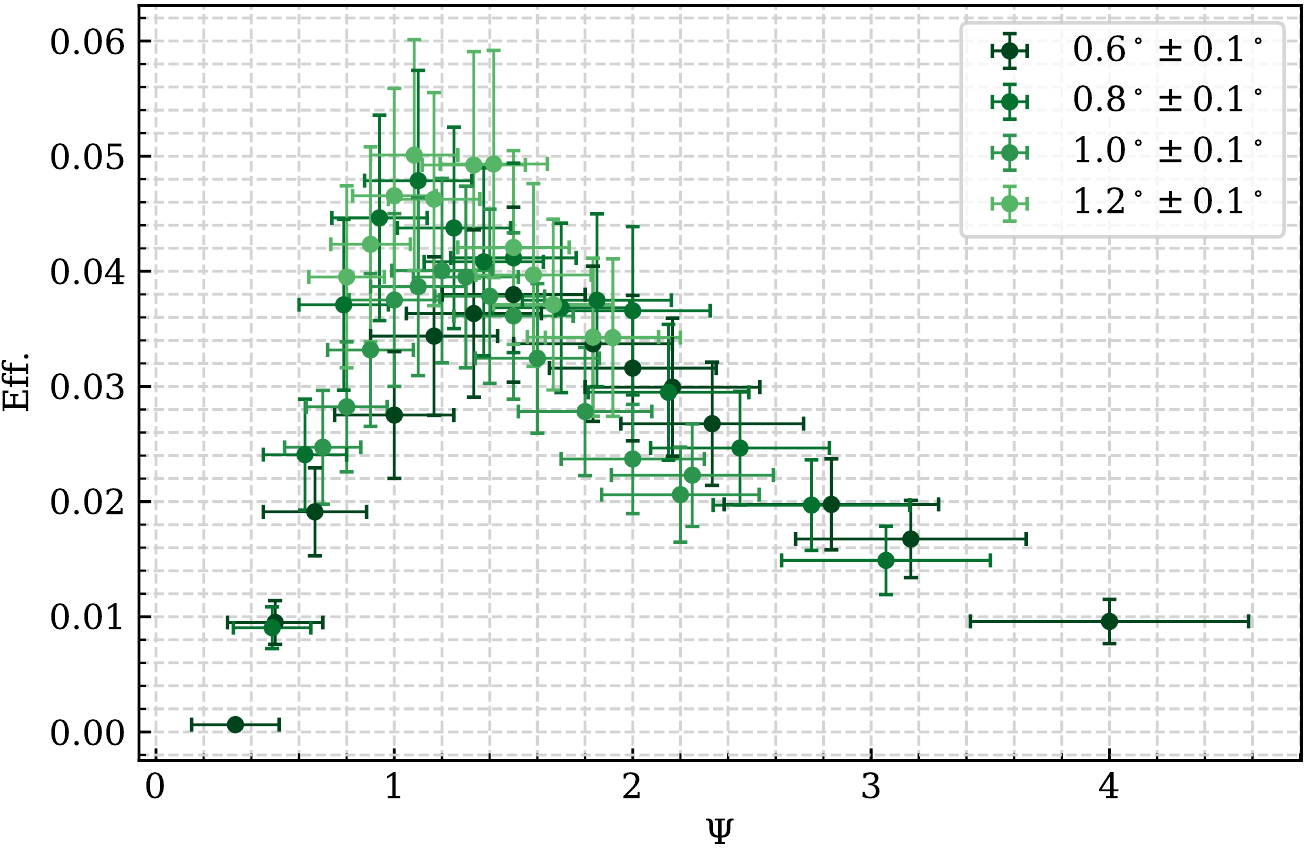} 
\includegraphics[width=1.\textwidth]{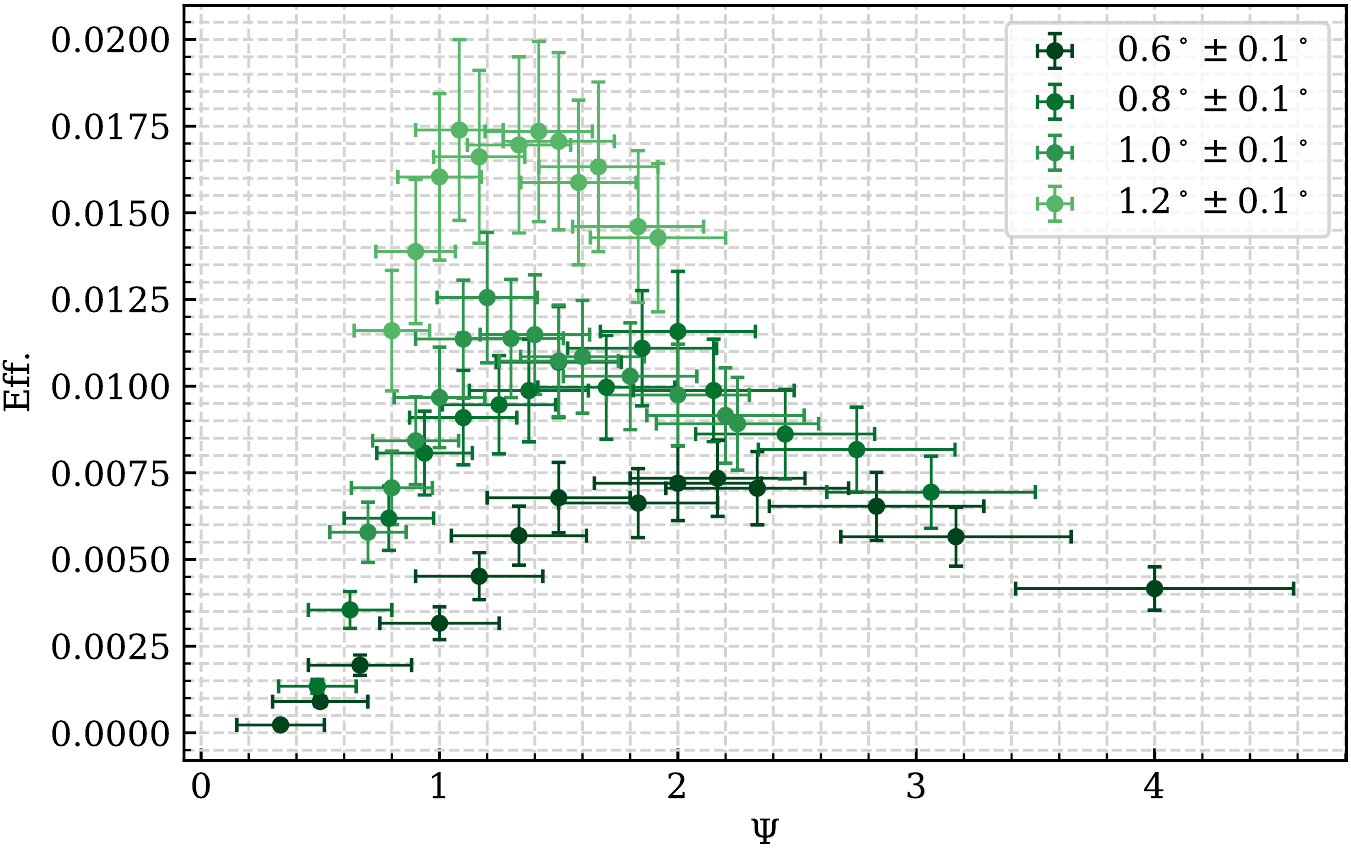}
\caption{Scattering efficiencies of the low-energy data set as a function of the scattering angle $\Psi=\theta/\theta_0$, for the incident angels of \ang{0.6}, \ang{0.8}, \ang{1.0}, and \ang{1.2}, for the on-axis (\textit{top panel}) and off-axis (\textit{bottom panel}) configurations. Energy beam of 172$\pm$30 keV.}
\label{fig:LE_measure}
\end{figure}

\begin{figure}[htbp]
\centering
\includegraphics[width=1.\textwidth]{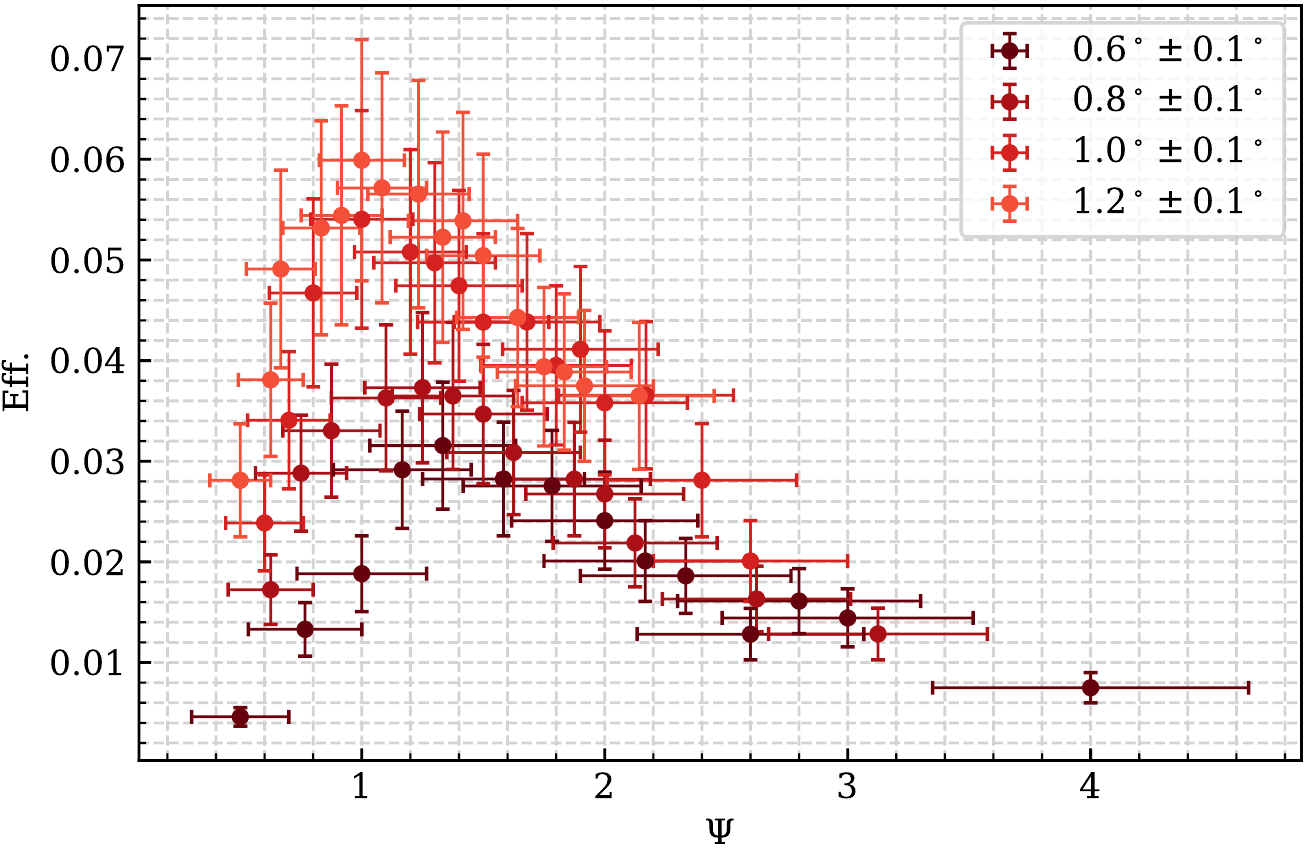}
\includegraphics[width=1.\textwidth]{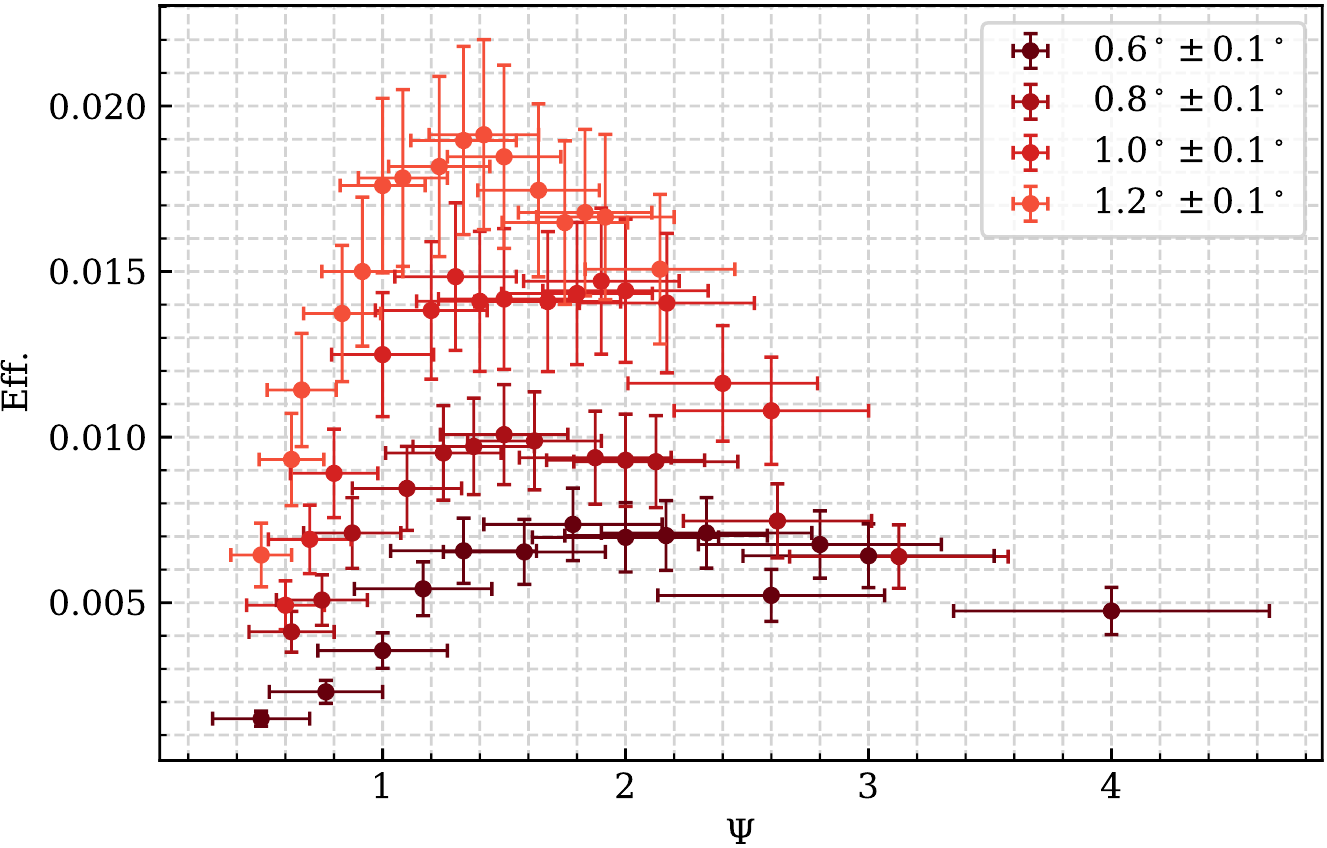}
\caption{Scattering efficiencies of the high-energy data set as a function of the scattering angle $\Psi$, for the incident angels of \ang{0.6}, \ang{0.8}, \ang{1.0}, and \ang{1.2}, for the on-axis (\textit{top panel}) and off-axis (\textit{bottom panel}) configurations. Energy beam of 471$\pm$25 keV.}
\label{fig:HE_measure}
\end{figure}

We measured the scattering efficiency at two different energies (high- and low-energy data sets, hereafter) and at four different incident angles: \ang{0.6}, \ang{0.8}, \ang{1.0}, and \ang{1.2}, both on-axis and  off-axis, the latter at an angle $\phi$ of about 2$^\circ$. Each data set consists of on-axis and off-axis scattering efficiencies. Results are shown in Figs.\,\ref{fig:LE_measure} and \ref{fig:HE_measure}, where the scattering efficiencies have been multiplied by the square of the incident angle \citep[as in][]{Amato2020} and are displayed as a function of the scattering angle divided by the incident one, i.e., $\Psi=\theta/\theta_0$.

For the high-energy data set (Fig.~\ref{fig:HE_measure}) we used a beam at $\sim$\SI{590}{\keV} from the accelerator, which was degraded by the Al foil down to 471$\pm$25~\si{\keV}. This energy value was chosen mainly for purposes of comparison with the previous measurements on the \erosita\ mirror sample \citep[][cfr. Fig.~\ref{fig:comparison_SPO_erosita}]{Diebold,SPIE2017}. The rationale behind the low-energy value can be found in the work of \citet{Lotti2017}, who showed that the highest transmission efficiency of soft protons is observed for those protons impacting the mirrors with 40-60 keV, for both instruments on board of \athena. Hence, it is crucial to investigate the scattering of soft protons at energies  around and below 100 keV. Unfortunately, the present setup could not reach such low energies, limiting us to a proton beam with an energy of $\sim$340~\si{\keV} at the exit of the accelerator, degraded to 172$\pm$30~\si{\keV} by the Al foil. In both cases, the values of the incident energies were determined by simulations with the software TRIM\footnote{The TRIM code is one of the SRIM (Stopping and Range of Ions in Solids) group of programs, available at \url{http://www.srim.org/index.htm\#HOMETOP}.} \citep[TRansport of Ions in Matter,][]{Ziegler2010}, already validated in \cite{Diebold}. 

As expected, the on-axis scattering efficiencies peak at the specular angle ($\psi\simeq1$) and are consistent with each other within the uncertainties. However, a higher spread is observed for the high-energy on-axis data set (Fig.\,\ref{fig:HE_measure}, top panel), with efficiencies ranging from 0.03 to 0.07 at the peak of the distribution. Also the off-axis data show a significant spread, which is  expected in this case. 

Overall, the maximum scattering efficiency values are $\sim$0.07 and $\sim$0.02 for the on-axis and off-axis configurations, respectively, with the low-energy data set showing slightly smaller efficiencies than the high-energy one.

\subsection{Comparison with the \textit{eROSITA} measurements}

\begin{figure}[!htbp]
\centering
\includegraphics[width=.48\textwidth]{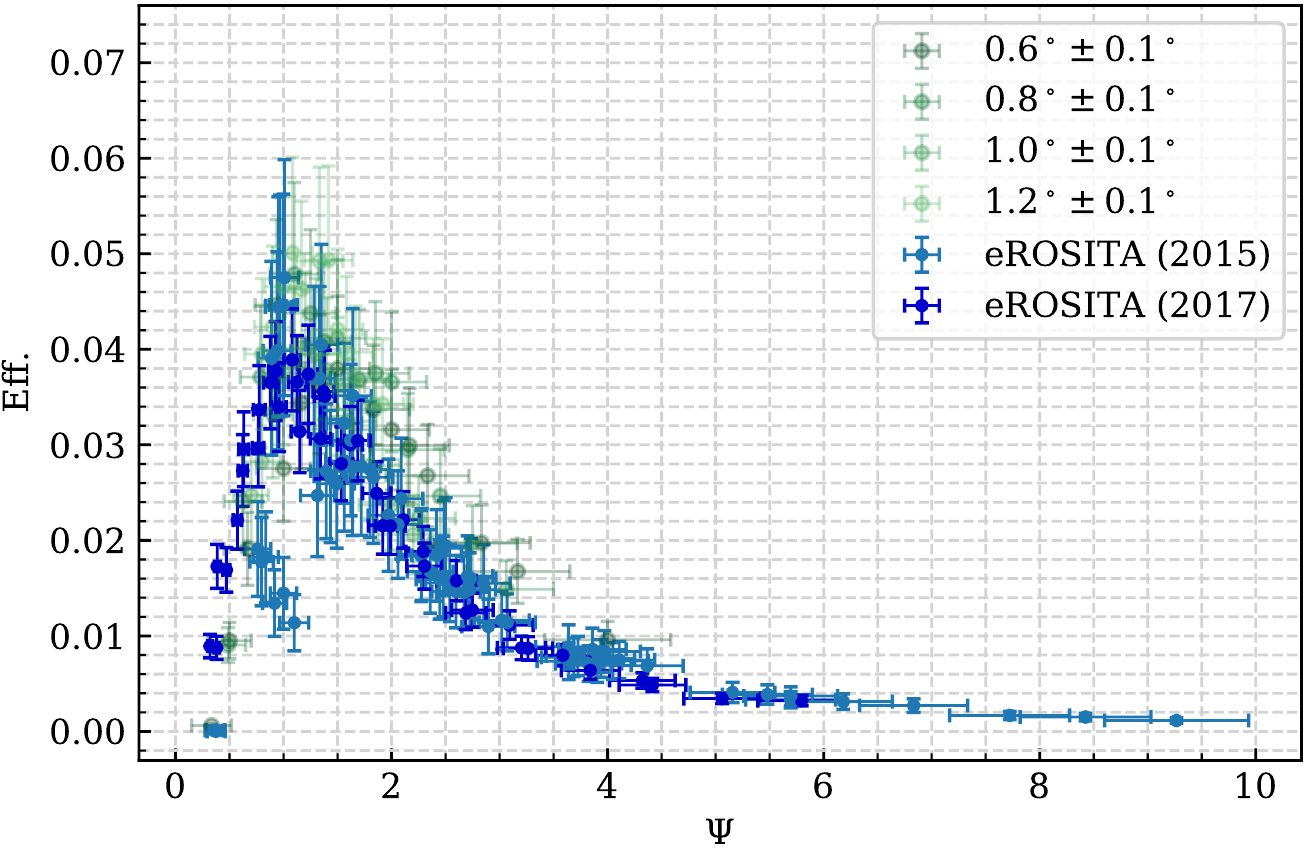} 
\includegraphics[width=.48\textwidth]{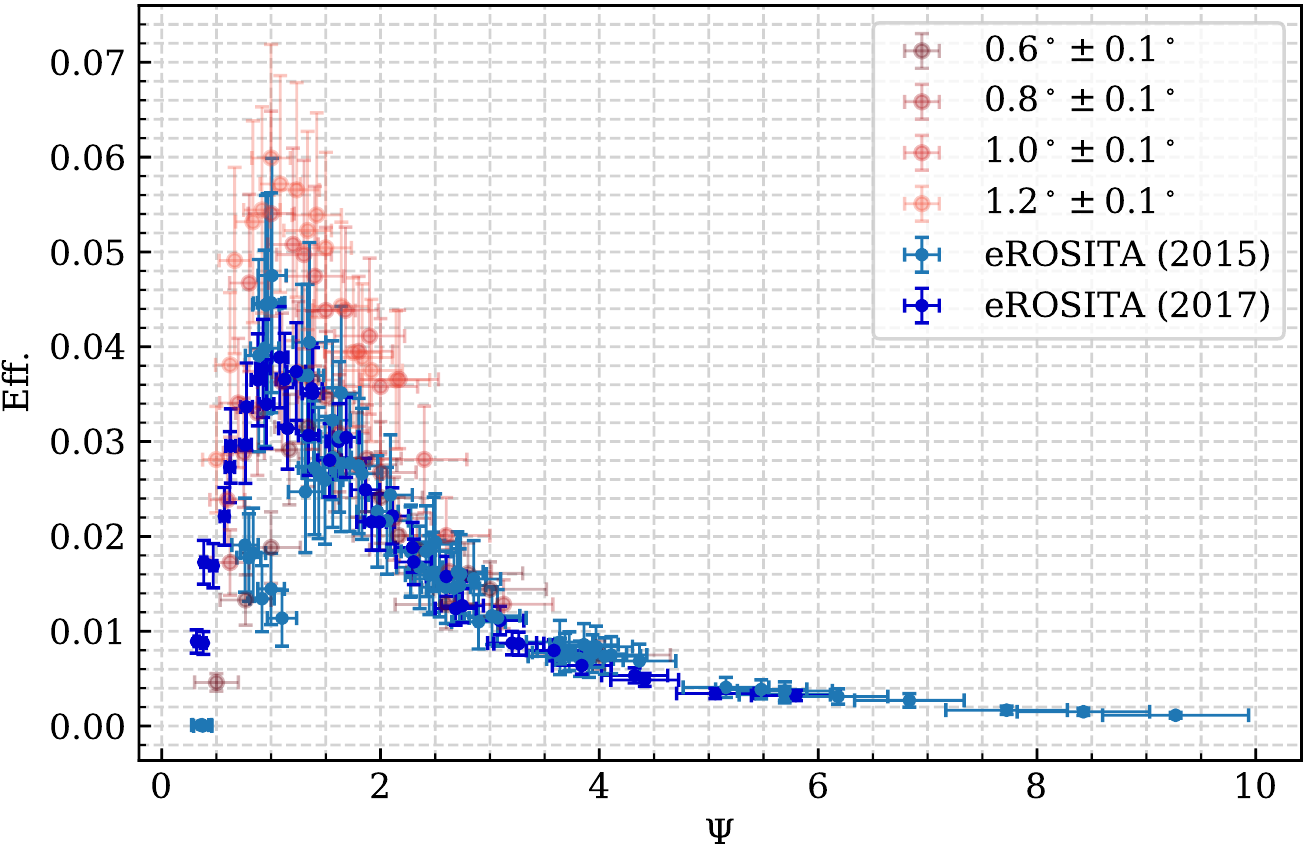}
\includegraphics[width=.48\textwidth]{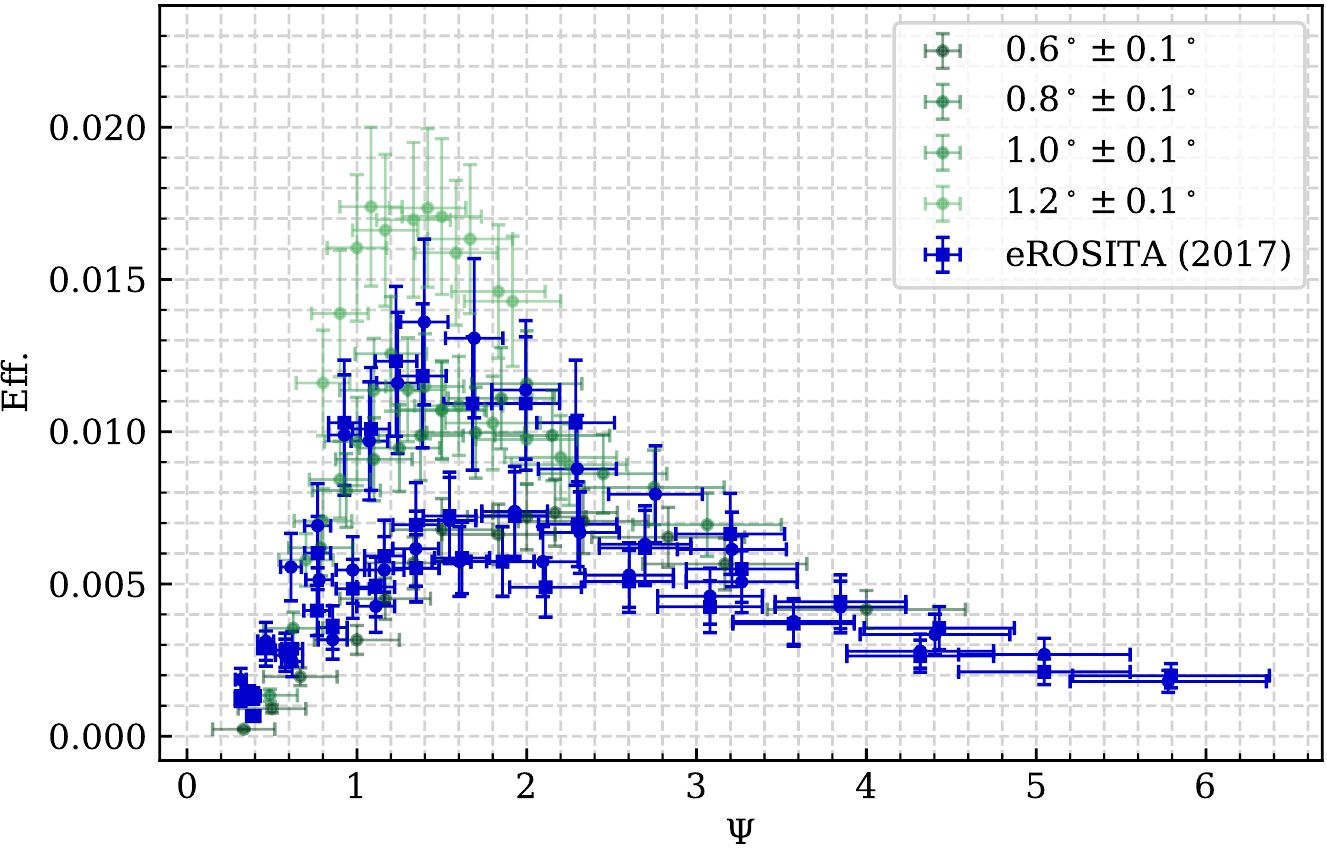}
\includegraphics[width=.48\textwidth]{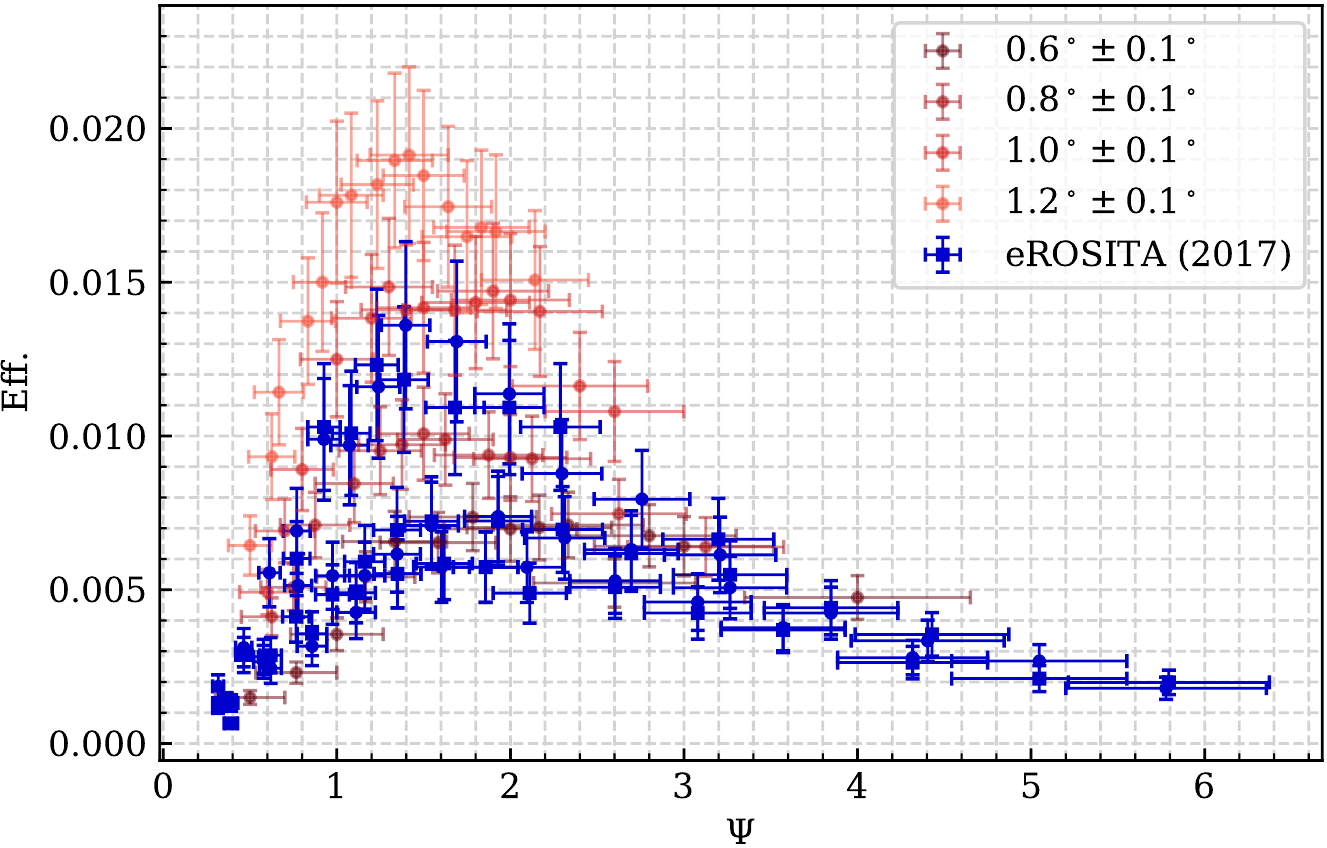}
\caption{Comparison of the \erosita\ scattering efficiencies (blue dots) with the SPO ones (green dots for the low-energy set and red dots for the high-energy set), for the on-axis (top panels) and off-axis (bottom panels) data.}
\label{fig:comparison_SPO_erosita}
\end{figure}

Fig.\,\ref{fig:comparison_SPO_erosita} shows the \erosita\ measurements \citep[][]{Diebold,SPIE2017} overlapped to the SPO data, for both the energies and the ox-axis and off-axis configurations.
Though the SPO efficiencies are systematically higher than the \erosita\ data, they are consistent within the error bars.

\begin{figure}[htbp]
\centering
\includegraphics[width=0.48\textwidth]{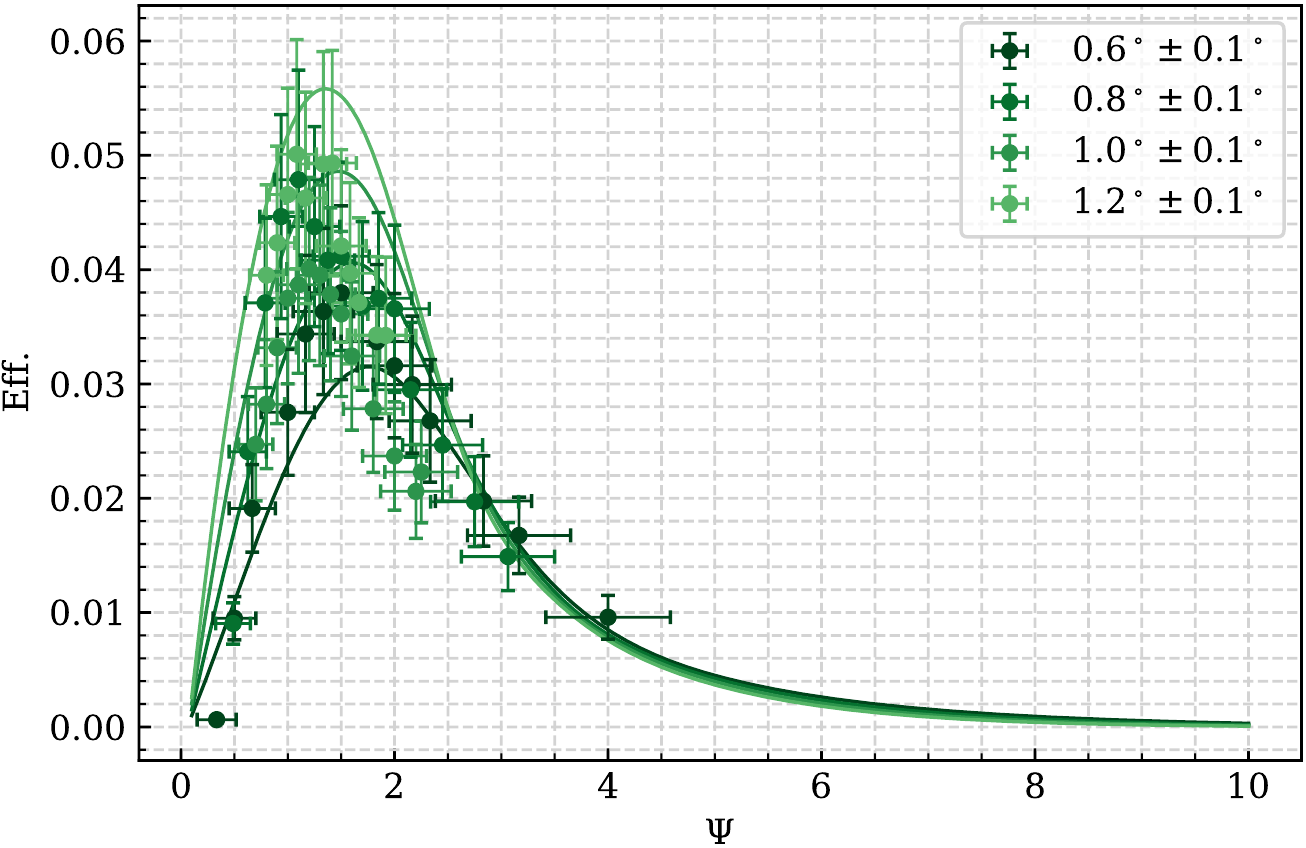}\quad
\includegraphics[width=0.48\textwidth]{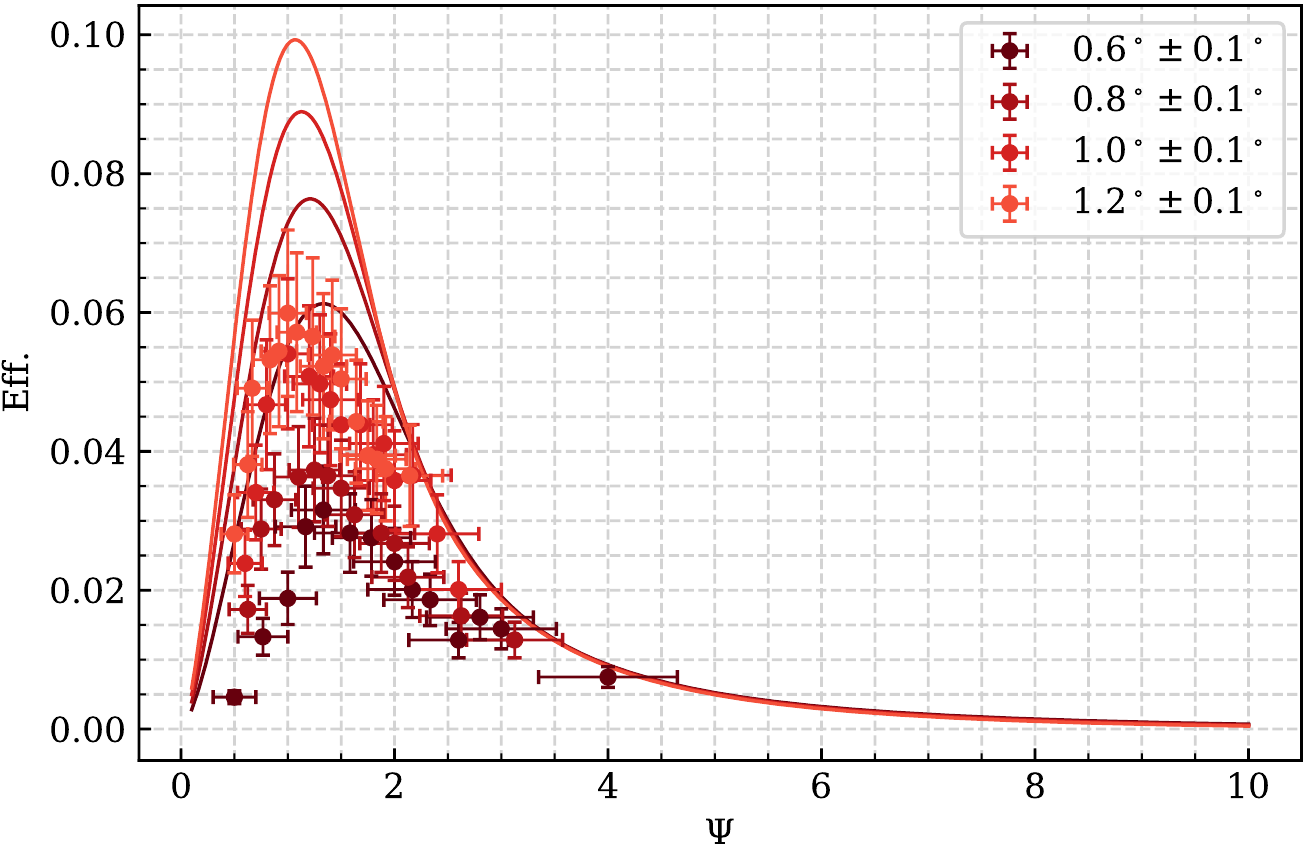}\quad
\includegraphics[width=0.48\textwidth]{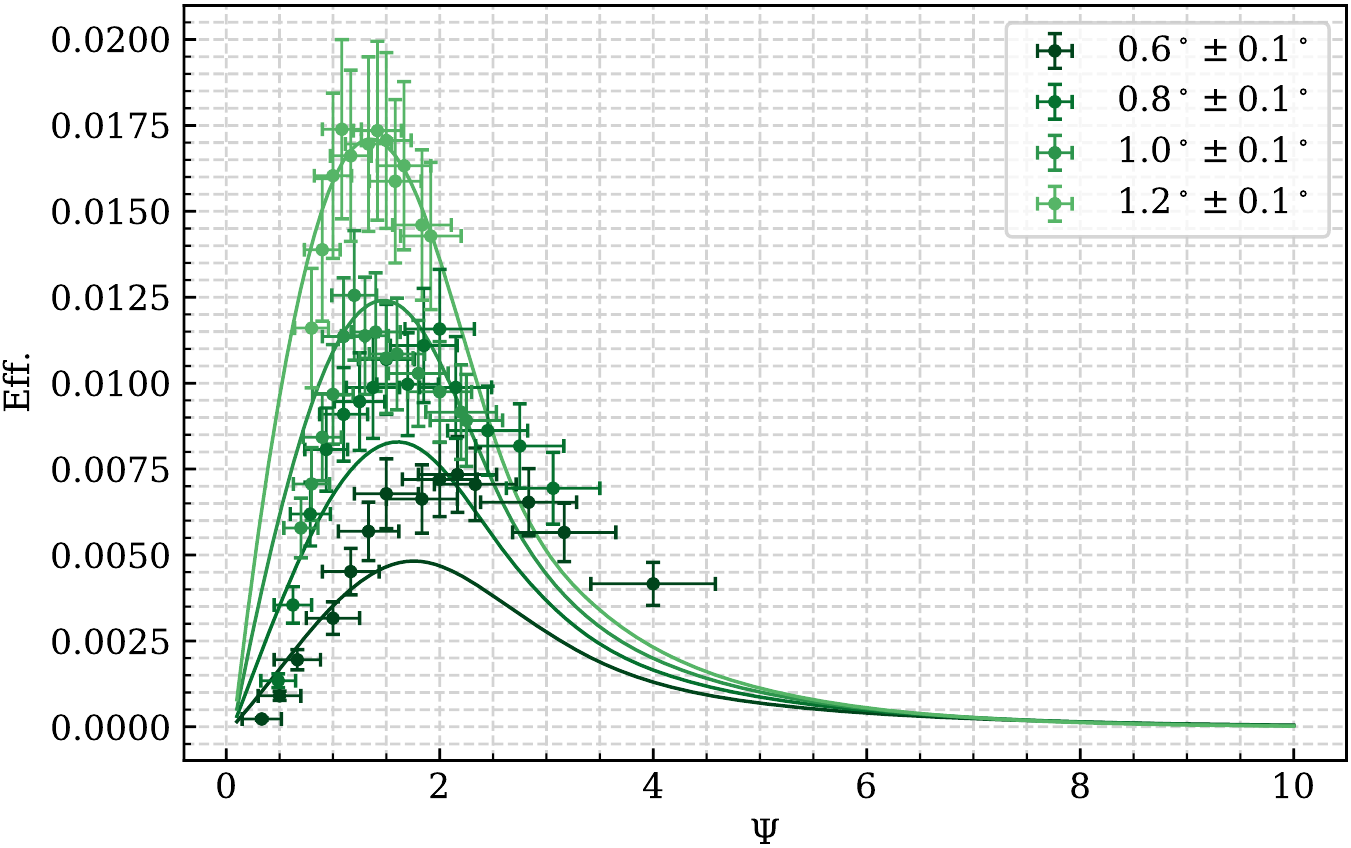}\quad
\includegraphics[width=0.48\textwidth]{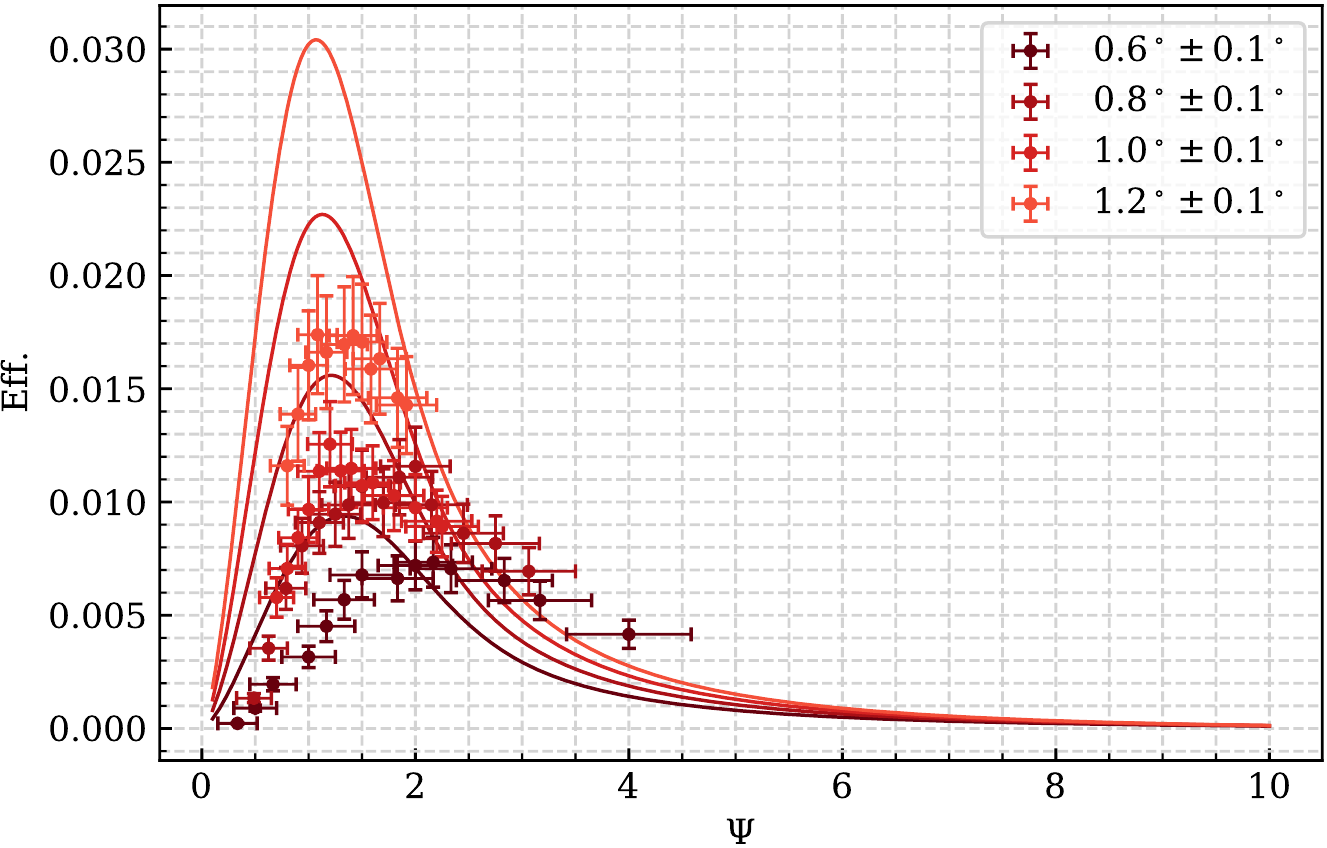}
\caption{Comparison between the experimental scattering efficiency of SPO (points) and the semi-empirical model developed from \erosita\ data (solid line), for the low-energy (green) and high-energy (red) data sets and for the on-axis (top panels) and off-axis (bottom panels) measurements. }
\label{fig:SPO_oldmodel}
\end{figure}

Due to this consistency, we applied the semi-empirical model proposed in \cite{Amato2020} to the scattering efficiencies of SPO, as shown in Fig.\,\ref{fig:SPO_oldmodel}. 

Overall, the model well reproduces the scattering efficiency of the low-energy data set, but overestimates the efficiency of the high-energy data set by a factor of $\sim$1.5 times. Nonetheless, it has to be borne in mind that at this stage we simply overlapped the semi-empirical model developed for the \erosita{} to the new SPO experimental data. A more accurate model, specific for SPO, can be obtained by fitting the data with the formula of \cite{Remizovich} in non-elastic approximation, as in \cite{Amato2020}, provided that energy loss measurements are retrieved from the raw data.

\begin{figure}[ht]
\centering
\includegraphics[width=0.48\textwidth]{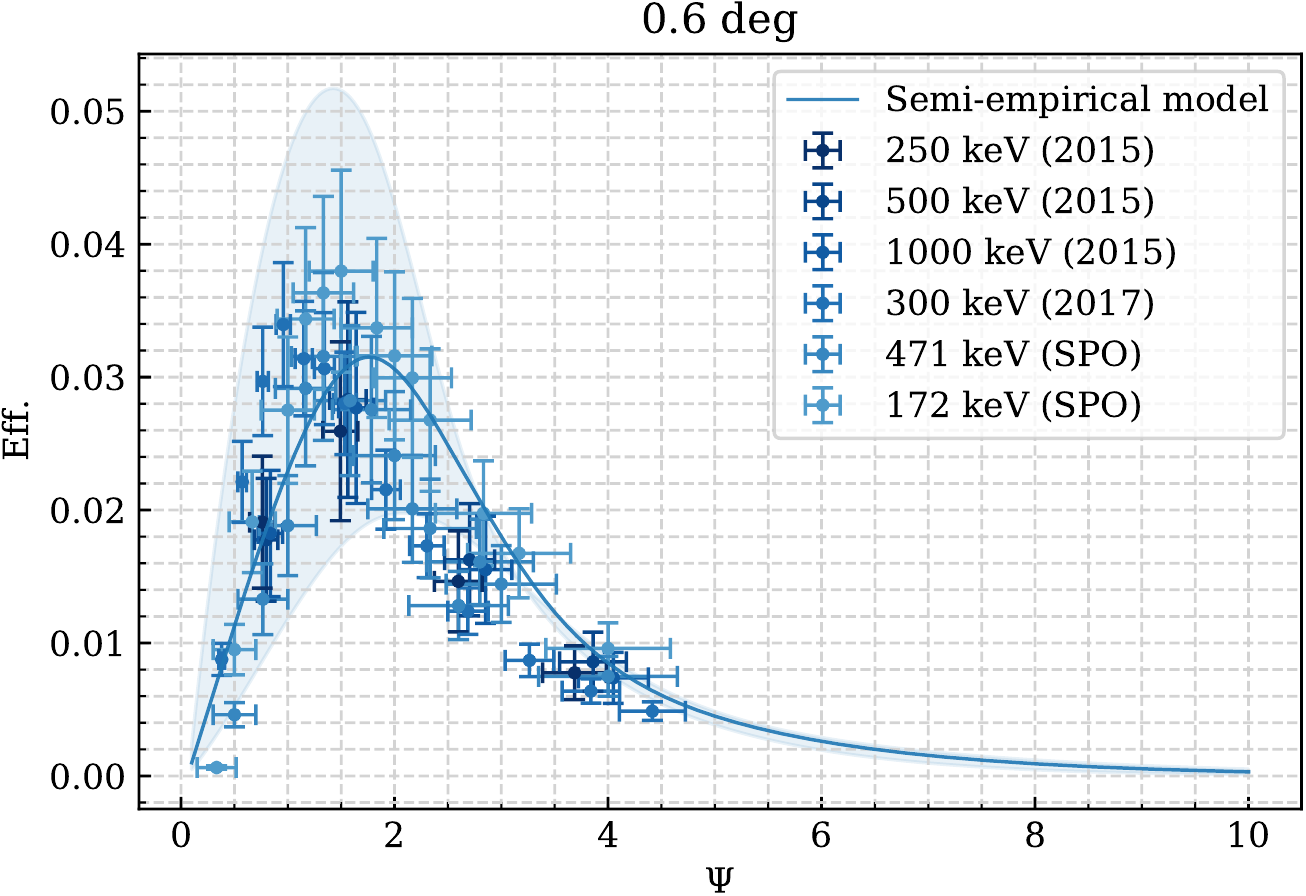} \quad
\includegraphics[width=0.48\textwidth]{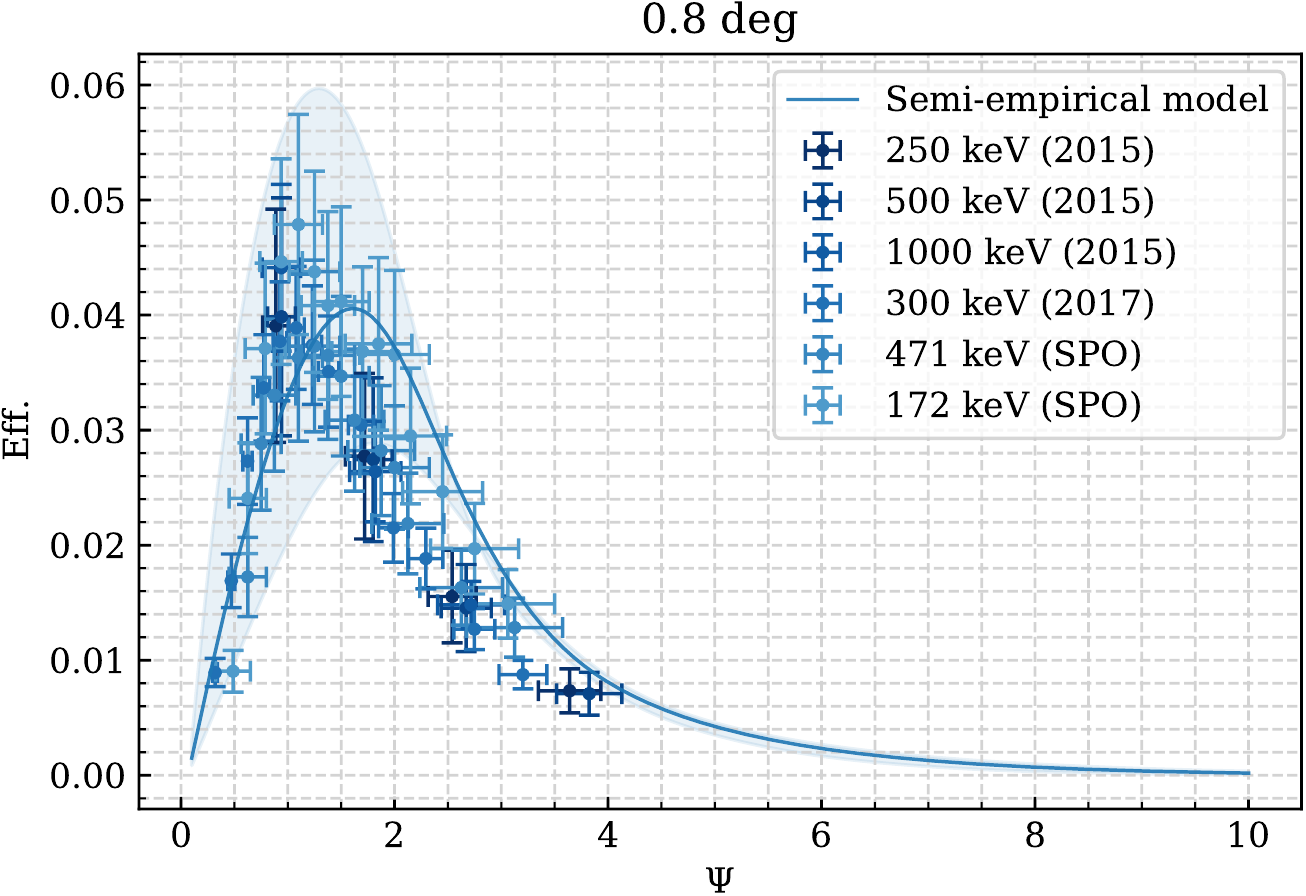} \quad
\includegraphics[width=0.48\textwidth]{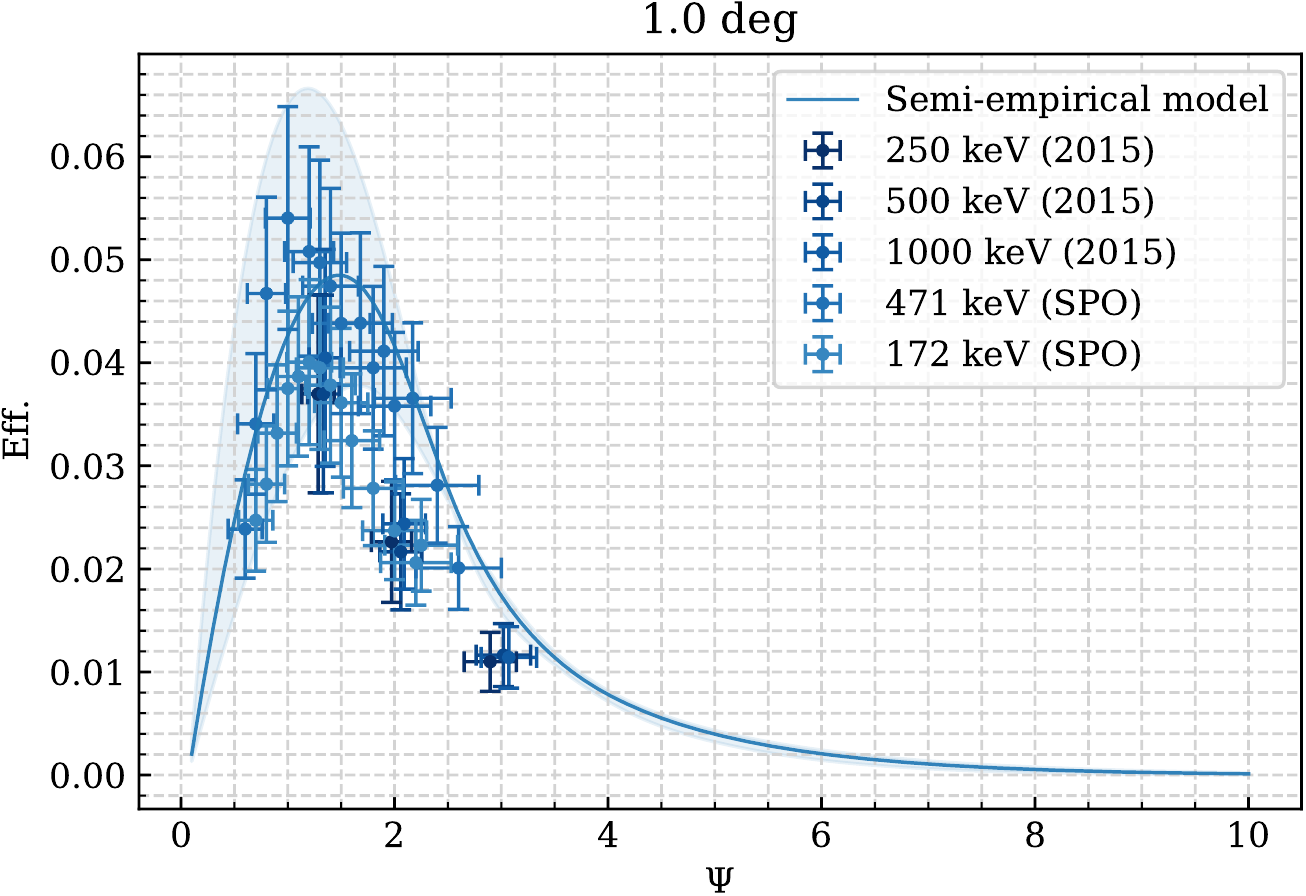} \quad
\includegraphics[width=0.48\textwidth]{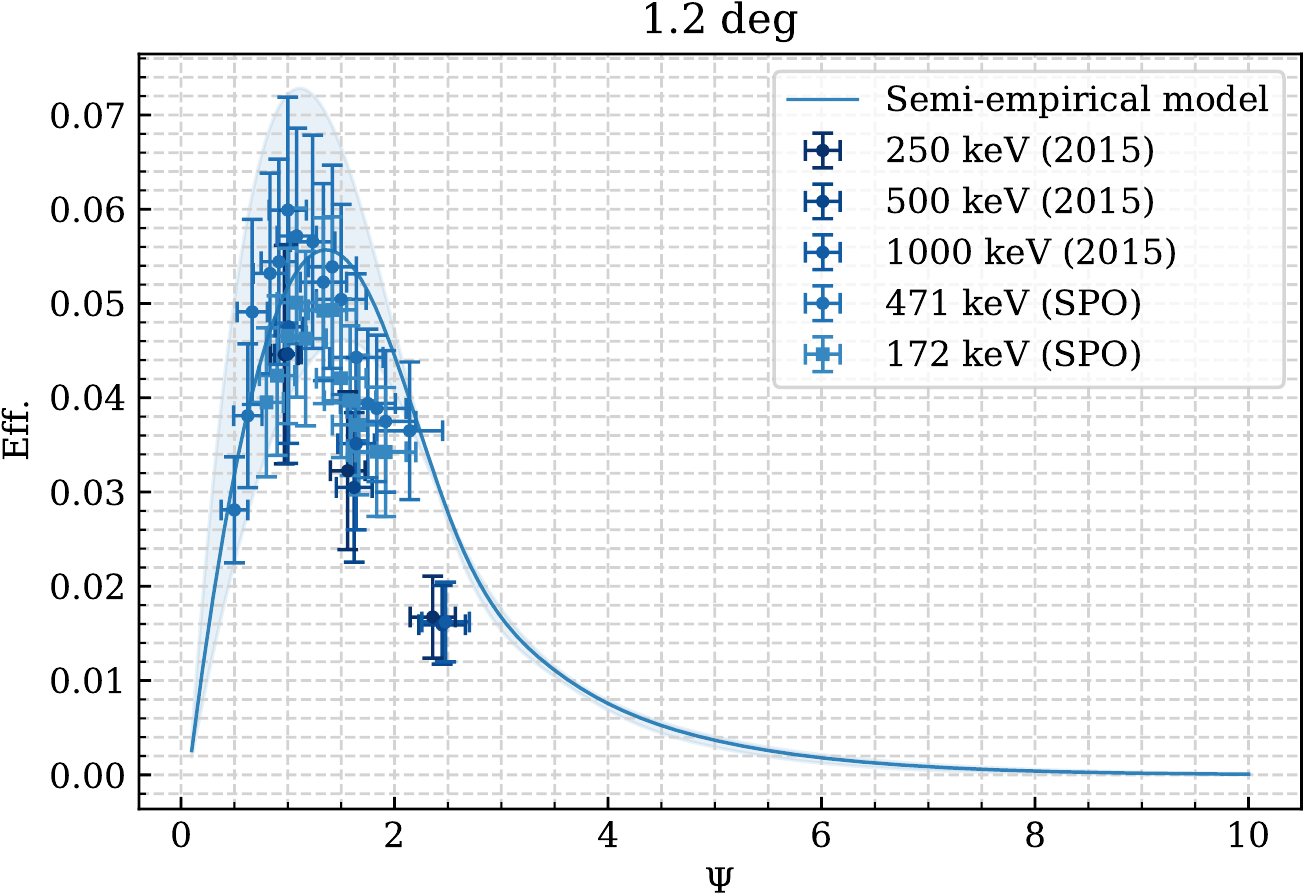}
\caption{Scattering efficiencies from \erosita{} mirror and the SPO samples, grouped by different incident angles, irrespective of the energy of the impinging beam. The solid line stand for the semi-empirical scattering model, with the error on the resulting efficiency given by the green area.}
\label{fig:data_SPO_sort_by_angle}
\end{figure}

Lastly, we group the efficiency values from the two data sets by the incident angle, irrespective of the energy of the incident beam. Fig.\,\ref{fig:data_SPO_sort_by_angle} shows that the data are perfectly consistent with each other and with the old \erosita\ measurements when grouped by the incident angle, without accounting for the energy. Once again, the semi-empirical model derived for the \erosita\ data is overlapped with the data, resulting in a general acceptable agreement.

\section{Conclusions}
\label{section:conclusions}

Within the EXACRAD project, we measured for the first time the scattering efficiency of a single wafer of SPO hit by low-energy protons at grazing incidence. Measurements were performed at two different energies, of about \SI{470}{\keV} and \SI{170}{\keV}, and at four different incident angles, \ang{0.6}, \ang{0.8}, \ang{1.0}, and \ang{1.2}, both on-axis and at an off-axis angle of about 2$^\circ$.

Hereafter some major remarks:
\begin{itemize}
    \item the scattering efficiencies show the trend expected from \cite{Remizovich} and from the experimental data on \erosita\ \citep{Diebold,SPIE2017}. The on-axis data peak close to  the specular reflection, while the off-axis data show a peak shifted to higher $\Psi$; the off-axis data reach lower efficiencies than the on-axis ones; higher incident angles resulted in higher scattering efficiency; 
    \item the SPO data are generally consisted with the \erosita\ data, though the high-energy data set show a higher spread in efficiency;
    \item as for the \erosita\ data, the scattering efficiency very weakly depends on the energy of the incident beam;
    \item the semi-empirical model developed from \erosita\ experimental data is able to acceptably reproduce the low-energy data set, while it results in higher efficiencies for the high-energy data set. The same model can be improved specifically for SPO, with a direct fit of the experimental data, provided that energy loss measurements are retrieved from the raw data.
\end{itemize}

As stated above, the experimental configuration used for this experiment is not suited for measurements at energies below 100 keV, which are those expected to contribute the most to the background of \athena\ \citep{Lotti2018}. Therefore, some possible changes to the setup could be explored in the future to enable measurements at lower energy ranges.\\

The work here presented is only the first step towards a thorough estimation of the SP flux expected at the focal plane of \athena. Indeed, after being scattered by the optics, SP cross all the other elements along their path, and, in particular, the filters located in front of each detector. As showed in \citet{Lotti2018}, the filters not only reduce the energy of the protons, but also act as degraders, altering their original trajectories. Hence, a thorough prediction of the overall transmission energy of SP can be achieved only combining the processes of scattering from the optics, crossing of the filters and energy release within the detectors. The latter two phenomena can be investigated, for instance, by simulations, as done, e.g., by \citet{Fioretti2018}.
If the estimated soft proton flux should still result higher than the scientific requirement of $5\times10^{-4}$ cts s$^{-1}$ cm$^{-2}$ keV$^{-1}$, in the 2--10\,keV energy band, for 90\,\% of the observing time (cfr. Sect.~\ref{section:introduction}), then appropriate solution should be adopted. Currently, the hypothesis of a magnetic diverter specific for protons is under discussion and possible designs are under investigation.

\begin{acknowledgements}
The experiment was conducted under the ESA contract n°4000121062
/17/NL/LF. Authors acknowledge S. Molendi, S. Lotti, and P. Nieminen as coordinators of the EXACRAD project; M. Collon, for providing the SPO targets for the experiment; S. Renner and S. Rieckert from the IAAT mechanical workshop for contributing to the beamline setup; L. Schmidt and the colleagues of the Van der Graaff accelerator facility for the onsite help.

This version of the article has been accepted for publication, after peer review but is not the Version of Record and does not reflect post-acceptance improvements, or any corrections. The Version of Record is available online at: \url{http://dx.doi.org/10.1007/s10686-021-09806-9}.
\end{acknowledgements}

%\addcontentsline{toc}{section}{\numberline{}\bibname}
\bibliographystyle{spbasic}      % basic style, author-year citations

\begin{small}
\bibliography{biblio.bib}

\begin{thebibliography}{14}
\providecommand{\natexlab}[1]{#1}
\providecommand{\url}[1]{{#1}}
\providecommand{\urlprefix}{URL }
\expandafter\ifx\csname urlstyle\endcsname\relax
  \providecommand{\doi}[1]{DOI~\discretionary{}{}{}#1}\else
  \providecommand{\doi}{DOI~\discretionary{}{}{}\begingroup
  \urlstyle{rm}\Url}\fi
\providecommand{\eprint}[2][]{\url{#2}}

\bibitem[{{Amato} et~al.(2020){Amato}, {Mineo}, {D'A{\i}}, {Diebold},
  {Fioretti}, {Guzman}, {Lotti}, {Macculi}, {Molendi}, {Perinati}, {Tenzer},
  and {Santangelo}}]{Amato2020}
{Amato} R, {Mineo} T, {D'A{\i}} A, et~al. (2020) {Soft proton scattering at
  grazing incidence from X-ray mirrors: analysis of experimental data in the
  framework of the non-elastic approximation}. Experimental Astronomy
  49(3):115--140, \doi{10.1007/s10686-020-09657-w}, \eprint{2003.07295}

\bibitem[{{Diebold} et~al.(2015){Diebold}, {Tenzer}, {Perinati}, {Santangelo},
  {Freyberg}, {Friedrich}, and {Jochum}}]{Diebold}
{Diebold} S, {Tenzer} C, {Perinati} E, et~al. (2015) {Soft proton scattering
  efficiency measurements on x-ray mirror shells}. Experimental Astronomy
  39:343--365, \doi{10.1007/s10686-015-9451-4}, \eprint{1504.01024}

\bibitem[{{Diebold} et~al.(2017){Diebold}, {Hanschke}, {Perinati}, {Smith},
  {Tenzer}, {Santangelo}, and {Jochum}}]{SPIE2017}
{Diebold} S, {Hanschke} S, {Perinati} E, et~al. (2017) {Updates on experimental
  grazing angle soft proton scattering}. In: Society of Photo-Optical
  Instrumentation Engineers (SPIE) Conference Series, Society of Photo-Optical
  Instrumentation Engineers (SPIE) Conference Series, vol 10397, p 103970W,
  \doi{10.1117/12.2272930}

\bibitem[{{Fioretti} et~al.(2018){Fioretti}, {Bulgarelli}, {Molendi}, {Lotti},
  {Macculi}, {Barbera}, {Mineo}, {Piro}, {Cappi}, {Dadina}, {Meidinger}, {von
  Kienlin}, and {Rau}}]{Fioretti2018}
{Fioretti} V, {Bulgarelli} A, {Molendi} S, et~al. (2018) {Magnetic Shielding of
  Soft Protons in Future X-Ray Telescopes: The Case of the ATHENA Wide Field
  Imager}. Astrophysical Journal 867(1):9, \doi{10.3847/1538-4357/aade99},
  \eprint{1808.09431}

\bibitem[{{Ghizzardi} et~al.(2017){Ghizzardi}, {Marelli}, {Salvetti},
  {Gastaldello}, {Molendi}, {De Luca}, {Moretti}, {Rossetti}, and
  {Tiengo}}]{XMMbkg-3}
{Ghizzardi} S, {Marelli} M, {Salvetti} D, et~al. (2017) {A systematic analysis
  of the XMM-Newton background: III. Impact of the magnetospheric environment}.
  Experimental Astronomy 44:273--285, \doi{10.1007/s10686-017-9554-1},
  \eprint{1705.04173}

\bibitem[{{Jansen} et~al.(2001){Jansen}, {Lumb}, {Altieri}, {Clavel}, {Ehle},
  {Erd}, {Gabriel}, {Guainazzi}, {Gondoin}, {Much}, {Munoz}, {Santos},
  {Schartel}, {Texier}, and {Vacanti}}]{Jansen2001}
{Jansen} F, {Lumb} D, {Altieri} B, et~al. (2001) {XMM-Newton observatory. I.
  The spacecraft and operations}. Astronomy and Astrophysics 365:L1--L6,
  \doi{10.1051/0004-6361:20000036}

\bibitem[{{Lotti} et~al.(2017){Lotti}, {Mineo}, {Jacquey}, {Molendi},
  {D'Andrea}, {Macculi}, and {Piro}}]{Lotti2017}
{Lotti} S, {Mineo} T, {Jacquey} C, et~al. (2017) {The particle background of
  the X-IFU instrument}. Experimental Astronomy 44:371--385,
  \doi{10.1007/s10686-017-9538-1}, \eprint{1705.04076}

\bibitem[{{Lotti} et~al.(2018){Lotti}, {Mineo}, {Jacquey}, {Laurenza},
  {Fioretti}, {Minervini}, {Santin}, {Molendi}, {Alberti}, {Dondero},
  {Mantero}, {Ivanchencko}, {Macculi}, and {Piro}}]{Lotti2018}
{Lotti} S, {Mineo} T, {Jacquey} C, et~al. (2018) {Soft proton flux on ATHENA
  focal plane and its impact on the magnetic diverter design}. Experimental
  Astronomy 45(3):411--428, \doi{10.1007/s10686-018-9599-9}

\bibitem[{{Nandra} et~al.(2013){Nandra}, {Barret}, {Barcons}, {Fabian}, {den
  Herder}, {Piro}, {Watson}, {Adami}, {Aird}, {Afonso}, and
  et~al.}]{Nandra2013}
{Nandra} K, {Barret} D, {Barcons} X, et~al. (2013) {The Hot and Energetic
  Universe: A White Paper presenting the science theme motivating the Athena+
  mission}. arXiv e-prints \eprint{1306.2307}

\bibitem[{Rasmussen et~al.(1999)Rasmussen, Chervinsky, and
  Golovchenkor}]{Rasmussen}
Rasmussen A, Chervinsky J, Golovchenkor J (1999) {Proton scattering off of XMM
  optics: XMM mirror and RGS grating sample}. Rgs-col-cal-99009, Columbia
  Astrophysics Laboratory

\bibitem[{{Remizovich} et~al.(1980){Remizovich}, {Ryazanov}, and
  {Tilinin}}]{Remizovich}
{Remizovich} VS, {Ryazanov} MI, {Tilinin} IS (1980) {Energy and angular
  distributions of particles reflected in glancing incidence of a beam of ions
  on the surface of a material}. Soviet Journal of Experimental and Theoretical
  Physics 52:225

\bibitem[{{Weisskopf} et~al.(2000){Weisskopf}, {Tananbaum}, {Van Speybroeck},
  and {O'Dell}}]{Weisskopf2000}
{Weisskopf} MC, {Tananbaum} HD, {Van Speybroeck} LP, {O'Dell} SL (2000)
  {Chandra X-ray Observatory (CXO): overview}. In: {Truemper} JE, {Aschenbach}
  B (eds) X-Ray Optics, Instruments, and Missions III, Proceedings of the SPIE,
  vol 4012, pp 2--16, \doi{10.1117/12.391545}, \eprint{astro-ph/0004127}

\bibitem[{{Willingale} et~al.(2013){Willingale}, {Pareschi}, {Christensen}, and
  {den Herder}}]{Willingale2013}
{Willingale} R, {Pareschi} G, {Christensen} F, {den Herder} JW (2013) {The Hot
  and Energetic Universe: The Optical Design of the Athena+ Mirror}. arXiv
  e-prints arXiv:1307.1709, \eprint{1307.1709}

\bibitem[{{Ziegler} et~al.(2010){Ziegler}, {Ziegler}, and
  {Biersack}}]{Ziegler2010}
{Ziegler} JF, {Ziegler} MD, {Biersack} JP (2010) {SRIM - The stopping and range
  of ions in matter (2010)}. Nuclear Instruments and Methods in Physics
  Research B 268(11-12):1818--1823, \doi{10.1016/j.nimb.2010.02.091}

\end{thebibliography}
\end{small}

%\clearpage
%\appendix
%\input{appendix}

\end{document}